\documentclass[aps,pre,twocolumn,superscriptaddress]{revtex4-1}
\usepackage[utf8]{inputenc}
\usepackage{amsmath}
\usepackage{amssymb}
\usepackage{epsfig}
\usepackage{graphicx,url}
\usepackage{bm}
\usepackage{mathrsfs}
\usepackage{tikz}
\usepackage{comment}

\begin{document}

\title{Optimal number of agents in a collective search and when to launch them}

\author{Hugues Meyer}
\affiliation{School of Physics and Astronomy, University of Nottingham, Nottingham NG7 2RD, United Kingdom}
\affiliation{Department of Theoretical Physics \& Center for Biophysics, Saarland University, 66123 Saarbrücken, Germany}
\author{Heiko Rieger}
\affiliation{Department of Theoretical Physics \& Center for Biophysics, Saarland University, 66123 Saarbrücken, Germany}

\date{\today}

\begin{abstract}
Search processes often involve multiple agents that collectively search a randomly located target. While increasing the number of agents usually decreases the time at which the first agent finds the target, it also requires resources to create and sustain more agents. In this manuscript, we raise the question of the optimal timing for launching multiple agents in a search in order to reach the best compromise between minimizing the overall search time and minimizing the costs associated with launching and sustaining agents. After introducing a general formalism for independent agents in which we allow them to be launched at arbitrary times, we investigate by means of analytical calculations and numerical optimization the optimal launch strategies to optimize the quantiles of the search cost and its mean. Finally, we compare our results with the case of stochastic resetting and study the conditions under which it is preferable to launch new searchers rather than resetting the first one to its initial position.     
\end{abstract}

\maketitle

\section{Introduction}

The term search processes encompasses any phenomenon in which the encounter of agents with a target is important \cite{grebenkov2024target}. They are encountered in a wide variety of systems across many length and time scales. This includes chemical reaction kinetics \cite{collins1949diffusion, rice1985diffusion},  biological processes at the molecular scale such as macromolecules searching for reactive sites to trigger biochemical reactions \cite{kolomeisky2011physics}, or at the cellular level in the immune search for pathogens \cite{harlin2009chemokine, krummel2016t, galeano2020cytotoxic}, ecological phenomena such as animal foraging and hunting \cite{hassell1978foraging, bartumeus2009optimal, sumpter2010collective, viswanathan2011physics, pyke2019animal}, or even searches by artificial agents such as robot swarms used in rescue missions  \cite{senanayake2016search, drew2021multi}.

Although stochastic search processes differ in nature, they share common features that can be treated equally using the tools of statistical physics. In particular, they share the need of being optimized in a certain way raising the question: how should one tune the free parameters of the process in order to make the search as efficient as possible? In a very large majority of studies, the search efficiency is fully defined in terms of the search time, i.e. the time required by the searchers to find and catch the target. Typically, the inverse of the mean first-passage time is used as an estimator of the search efficiency, which one attempts to maximize. For individual agents evolving according to stochastic dynamics, a lot of work has been dedicated over the past years to identify search strategies, meaning parameter sets of the stochastic search process, that minimize the mean first-passage time. Various classes of processes have been investigated, among which intermittent or Lévy walks  \cite{viswanathan1996levy, viswanathan1999optimizing, oshanin2007intermittent, yang2009cuckoo, reynolds2009levy, benichou2011intermittent}, stochastic resetting \cite{kusmierz2014first, chechkin2018random, bressloff2020search, evans2020stochastic, pal2020search}, or non-markovian searches \cite{tejedor2012optimizing, meyer2021optimal, barbier2022self, klimek2022optimal, altshuler2023environmental}. However, it is only rarely that the search efficiency has been defined to account for other quantities than the search time, such as the energy deployed for the search to succeed or the amount of external resources required. 

More recently, some effort has been put into understanding and quantifying collective searches, i.e. identifying optimal strategies that multiple searchers should follow when they are looking for a common target \cite{romanczuk2009collective, martinez2013optimizing,  janosov2017group, kamimura2019group, surendran2019spatial}. Most studies focused on the MFPT of independent searchers \cite{mejia2011first, bernardi2022run, biroli2023critical},  but also a few instances for interacting searchers have been studied \cite{martinez2013optimizing, tani2014optimal, ro2023target, meyer2024collective}. In particular, similarly to individual searches, the search efficiency is almost exclusively defined inversely proportional to the overall search time. In addition, the number of searchers has rarely been considered as a parameter to optimize. However, while it is clear that for nearly all search processes the mean search time decreases with the number of searchers, adding more searchers may have a non-negligible cost in terms of the required resources. For a human search problem where one needs to pay agents and material resources to perform the search, one can very easily understand that it is not optimal to hire as many agents as possibly available, as it would have a huge financial cost. Similarly, in an immune response process, hiring more cells to find a pathogen requires a substantial amount of metabolic energy. This is precisely the motivation for the question that we are raising in this paper: Given the cost associated with launching and sustaining an agent in a collective search process, what is the optimal number of them and when to launch them?

To address this general question, the manuscript is organized as follows. In section II we formalize the problem and discuss the assumption on which our work relies. We also introduce a definition of the search cost and discuss its various contributions. Section III is dedicated to the minimization of the quantiles of the search cost, i.e. the probability that the cost does not reach too large values. We show that the whole problem can be solved for logarithmically convex SASP and we therefore derive the optimal strategies that minimize the quantiles of the search cost. We then apply our result to a specific example, namely an exponentially decaying single-agent target survival probabilities (SASP). Section IV deals with the mean search cost. We first treat the 2-searcher problem to gain an intuition of the problem, and show that the condition on the log-convexity of the SASP is still relevant. We then derive general analytic results on the optimal number of searchers to launch initially and later perform numerical optimization on a selection of test cases that combine relevant short- and long-time behavior of the SASP to identify the details of the optimal launch strategies. Section V presents results of numerical simulations on an example of search process and compares the data with the predictions made in the previous sections. Finally, in section VI we draw a comparison between our formalism and search processes subject to stochastic resetting, which have attracted a lot of attention in recent years.  We show with the canonical example of the one-dimensional diffusive search under which conditions launching new searchers is preferable to resetting the first searcher to its initial position. We conclude in section VII in which we discuss possible adaptations of our work to more complex problems.

\section{General formalism}
\begin{figure}
	\begin{center}
		\includegraphics[width=\linewidth]{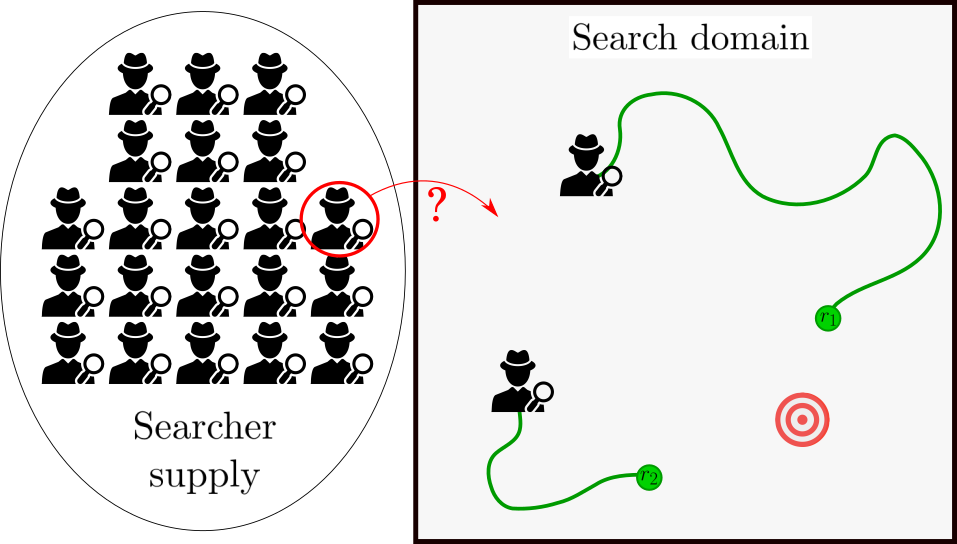}
		\caption{Sketch of the problem under study. A target is to be found in a search domain and we have a supply of searching agents at our disposal. Progressively, agents are launched into the search domain until the target has been found by one of them. The main question of the manuscript is then: given the probability for an individual agent to find the target at a certain time after being launched, how should one optimize the timing of the launches in order to minimize the toal cost of the process?}
		\label{fig:sketch_launch}
	\end{center}
\end{figure}

\subsection{Target survival probability}

The general problem that we consider is the following. A target is initially located at position at position $\mathbf{r}_T$ in a search domain that can be finite or infinite. We have at our disposal a supply of $N$ agents that are initially out of the search domain and ready to be launched into the search. Once in the search domain, each agent moves within it according to a certain stochastic dynamics. The $i$\textsuperscript{th} agent is launched at time $t_i \geq t_{i-1}$ -- referred to as the {\it launch time} -- and position $\mathbf{r}_i$. We set $t_1=0$ and we note $\Delta_i = t_i - t_{i-1}$ the $i$\textsuperscript{th} {\it launch interval}. The overall process terminates once one searcher finds the target. A sketch of the setup is depicted in figure \ref{fig:sketch_launch}. Depending on the search process, {\it finding} the target can be achieved whenever an agent comes in the vicinity of the target for the first time, while it may take several attempts for so-called {\it reactive} targets which need to interact with the agents in a certain way in order to trigger the initiation of a subsequent process, e.g. in chemical reactions. This distinction will be irrelevant in what follows and both scenarios can be treated by our formalism.

The individual processes can be arbitrary, provided that the single-agent survival probability (SASP) of the target $s_i(T,\mathbf{r}_i,\mathbf{r}_T)$ associated with the searcher $i$, i.e. the probability that the searcher $i$ has not found the target until time $t$, is well defined. Note that $s_i$ could be different for each searcher if their dynamics is not identical. Because the searchers are independent, the overall survival probability of the target is given by
\begin{equation}
S(T, \left\{t_i , \mathbf{r}_i \right\}, \mathbf{r}_T) = \prod_{k=1}^{n} s_k(T-t_k, \mathbf{r}_k, \mathbf{r}_T)
\end{equation}
for $t_{n}\leq T < t_{n+1}$. Defining $\varrho_T(\mathbf{r}_T)$ and $\varrho_S(\mathbf{r}_1,\cdots,\mathbf{r}_N)$ the probability distributions for the target position $\mathbf{r}_T$ and initial coordinates of each searcher $\mathbf{r}_1,...,\mathbf{r}_N$ respectively, the averaged survival probability is defined as 
\begin{widetext}
	\begin{equation}
	\bar{S}(T, \left\{t_i \right\}) =  \int d\mathbf{r}_T \int d\mathbf{r}_1 \cdots  \int d\mathbf{r}_N \varrho_T(\mathbf{r}_T)\varrho_S(\left\{\mathbf{r}_i \right\}) S(T, \left\{t_i , \mathbf{r}_i \right\}, \mathbf{r}_T) 
	\end{equation}
\end{widetext}
In order to proceed we now assume that (i) the single-agent survival probabilities are identical i.e. $s_k=s$ (ii) searchers launch positions are independent and identically distributed, i.e. $\varrho_S(\mathbf{r}_1,\cdots,\mathbf{r}_N) = \prod_{k=1}^{N}\rho_s(\mathbf{r}_k)$ . With this we find
\begin{widetext}
	\begin{equation}
	\label{eq:Survival}
	\bar{S}(T, \left\{t_i \right\}) = \int d\mathbf{r}_T  \varrho_T(\mathbf{r}_T) \left( \prod_{k=1}^{n} \int d\mathbf{r}_k \rho_s(\mathbf{r}_k) s(T-t_k, \mathbf{r}_k, \mathbf{r}_T) \right)
	\end{equation}
\end{widetext}
again for $t_n \leq T < t_{n+1}$. From here, a number of search processes are such that equation (\ref{eq:Survival}) can be simplified into the form  
\begin{equation}
\label{eq:joint_proba}
\bar{S}(T, \left\{t_i \right\}) =  \prod_{k=1}^{n} \bar{s}(T-t_k) 
\end{equation}
This includes two main classes of processes. First for a target fixed at a deterministic position $\mathbf{r}_T^0$, i.e.  $ \varrho_T(\mathbf{r}_T) =  \delta(\mathbf{r}_T-\mathbf{r}_T^0)$. In this case we have $\bar{s}(T-t_k) = \int d\mathbf{r}_k \rho_s(\mathbf{r}_k) s(T-t_k, \mathbf{r}_k, \mathbf{r}_T^0) $. Second, a finite translationally invariant search domain $\mathcal{V}$ with periodic boundary conditions such that $\bar{s}$ depends on $\mathbf{r}_k$ and  $\mathbf{r}_T$ through $\mathbf{r}_k- \mathbf{r}_T$, and where agents are launched homogeneously in the domain, i.e. $\rho_s(\mathbf{r}_k) = 1/V$ where $V$ is the volume of the search domain. Then we obtain  
\begin{align}
\int_\mathcal{V} d\mathbf{r}_k  \rho_s(\mathbf{r}_k)& s(T-t_k, \mathbf{r}_k-\mathbf{r}_T) \nonumber\\
&=\frac{1}{V} \int_{\mathcal{V}+ \mathbf{r}_T} d\mathbf{r}_k s(T-t_k, \mathbf{r}_k) \nonumber\\
&= \frac{1}{V} \int_{\mathcal{V}} d\mathbf{r}_k s(T-t_k, \mathbf{r}_k) \nonumber\\
&=  \bar{s}(T-t_k) 
\end{align}
where the second equality comes from periodicity. Note that these two examples are not necessarily the only possible ones that would result into equation (\ref{eq:joint_proba}).

From this point on, we will focus on situations where the joint probability $\bar{S}(t)$ can be written in the form of equation (\ref{eq:joint_proba}) such that our entire analysis will be made in terms of the averaged single-agent survival probability $\bar{s}(t)$. This will be the central quantity of the paper. Note, however, that this object can still depend on the initial distance between searchers and the target if their respective initial probabilities are delta-distributed. We finally make two assumptions on $\bar{s}(t)$: (i) $\lim_{t\to\infty} \bar{s}(t) = 0$, i.e. the probability for on agent to find the target eventually is 1, (ii) the function $\bar{s}: \mathbb{R}^+ \to [0,1]$ is bijective such that we can define an inverse function $\bar{s}^{-1}$ which obeys $\bar{s}^{-1}(\bar{s}(t)) = t$ for all $t>0$.

Finally, we emphasize that the launch times $t_i$ are chosen to be deterministic, i.e. they are not drawn from a certain probability distribution and do not require to be averaged over. This choice is motivated by the fact that search processes are often optimized when control parameters are not random. This is e.g. well-known for stochastic resetting where the resetting times are deterministic in optimal strategies \cite{chechkin2018random}, or in random search processes with memory where optimal transition rates are also found to be deterministic \cite{meyer2021optimal}.

\subsection{Search cost}

In order to identify optimal launch strategies, we first need to define a search cost, i.e. an objective function of the launch times $\left\{t_i \right\}$ that we will aim at minimizing. As mentioned in the introduction, we want to account not only for the overall search time but also for the resources required to launch each searcher and to sustain it. We thus aim at constructing a cost function that depends on the total number $\mathcal{N}$ of searchers launched before the target is found and the total time $\mathcal{T}$ spent by all searchers in the search. These quantities can be defined in terms of the total search time $T$, reading
\begin{align}
\mathcal{N} =& \sum_{i=1}^N \Theta(T-t_i) \\
\mathcal{T} =& \sum_{i=1}^N (T-t_i) \Theta(T-t_i) 
\end{align}
where $\Theta$ is the Heaviside step function. A relevant search cost $K$ should be a function of $T$, $\mathcal{N}$ and $\mathcal{T}$ where each of them should contribute independently. The simplest form that can be constructed is thus a weighted sum of three contributions, namely 
\begin{align}
\label{eq:def_cost}
K = J_T T + J_S \mathcal{N} + K_L  \mathcal{T}
\end{align}
The first term weighted by the {\it target cost rate} $J_T$ quantifies a cost associated with the presence of the target and can be interpreted as a rate of damage due to the presence of the target. The second term weighted by the {\it searcher sustaining rate}  $J_S$ quantifies the amount of resources required to sustain one searcher per unit time. Finally, the last term weighted by the {\it searcher launch cost} $K_L$ quantifies the amount of resources required to launch a searcher. The values of $J_T$, $J_S$ and $K_L$ are constant for each specific search process and reflect the relative importance of each contribution to the search cost. Of course, more complicated definitions of the search cost can be chosen where the dependence on $T$, $\mathcal{N}$ and $\mathcal{T}$ are non-linear, but this would be beyond the scope of this manuscript and is briefly discussed in the conclusion of this paper. For compactness, we introduce the normalized parameters $\gamma = J_S/J_T$ and $\kappa = K_L/J_T$ and set $J_T=1$ as our cost rate unit for the rest of the paper. 

Note that on the level of equation (\ref{eq:def_cost}), $K$ is a stochastic quantity as $T$, $\mathcal{N}$ and $\mathcal{T}$ differ from on trajectory to the other. Optimizing a search strategy implies to minimize a statistical estimator of $K$ with respect to the launch times $t_i$. In what follows, we will first focus on the optimization of the quantiles of the search cost, and then on the mean search cost. In both cases, we will show that some general results can be derived if one assumes the SASP to be logarithmically convex, which is actually observed in many search processes.

\section{Optimizing the quantiles}

One plausible criterion for optimizing the search strategy may be to minimize the probability that the search cost reaches too large values. This can be quantified through the $q$-quantile of the search cost distribution. Given the probability distribution $r_K(k)$ of the search cost, the $q$-quantile $\hat{k}_q$ is defined as the solution of 
\begin{equation}
\label{eq:def_quantile}
\int_0^{\hat{k}_q} r_K(k) dk = \frac{1}{q}
\end{equation}
In particular $\hat{k}_2$ is know as the median cost. Denoting $\mathcal{F}_K(k)$ the probability that the search cost is greater than $k$, equation (\ref{eq:def_quantile}) can also be written as $   \mathcal{F}_K(\hat{k}_q)  = 1 - q^{-1}$. Let us now note $z = 1 - q^{-1}$ and let $k_z$ be the solution of the equation 
\begin{equation}
\mathcal{F}_K(k_z)  = z
\end{equation}
With these notations, the probability that the search cost is larger than $k_z$ is $z$. Technically, we have $k_z = \hat{k}_{\frac{1}{1-z}}$, i.e. $k_z$ is the $(\frac{1}{1-z})$-quantile. Let us now show how to minimize it for a given value of $z$.\\

Before we proceed, we first introduce the function
\begin{align}
\zeta(t) =& \frac{\bar{s}'(t)}{\bar{s}(t)} 
\end{align}
which will be a useful object in the next paragraphs.

\subsection{Minimizing a single $q$-quantile}

\begin{figure*}
	\centering
	\includegraphics[width=.42\linewidth]{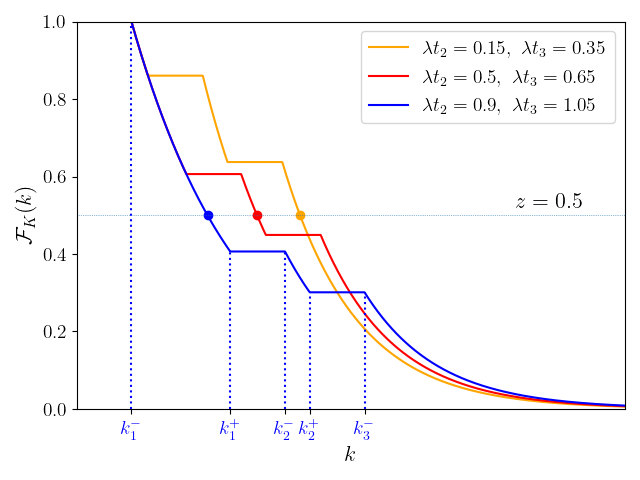}
	\includegraphics[width=.57\linewidth]{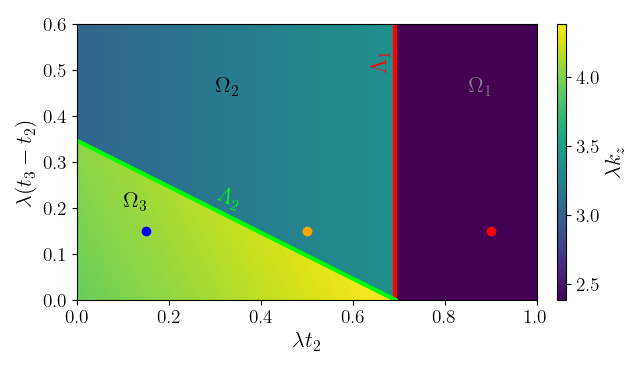}
	\caption{Quantile optimization for $N=3$ and $\bar{s}(t)=e^{-\lambda t}$ with $\gamma=1$ and $\kappa\lambda = 1$. $\underline{\text{Left}}$: We show $\mathcal{F}_K$ as a function of $k$ for various choices of $t_2$ and $t_3$. The dashed line indicates $z=0.5$, i.e. the median. Each of the curves is located in a different subdomain $\Omega_n$ in the $(t_2, t_3)$-plane. $\underline{\text{Right}}$: Median cost in the $(t_2, t_3)$-plane for the same values of $\kappa$ and $\gamma$. The boundaries $\Lambda_1$ and $\Lambda_2$ are shown as the solid lines and the three dots correspond to the three curves shown in the left panel. Here, the minimum of $k_z$ is located on $\Lambda_1$ defined as $\lambda t_2 = -\ln(z)$, where $k_z$ is constant. The optimal strategy consists here in launching the first searcher and wait a time at least longer than $-\ln(z)$ to launch the second and third searcher in order to minimize the quantile.}
	\label{fig:quantiles_N3}
\end{figure*}

Because the search cost $K$ is a piecewise affine function of the first-passage time $T$, we can straightforwardly relate $\mathcal{F}_K$ to $\bar{S}$ and in turn to $\bar{s}$ via 
\begin{widetext}
	\begin{equation}
	\label{eq:def_quantiles}
	\mathcal{F}_K(k) = \left\{ 
	\begin{tabular}{ll}
	$\prod_{i=1}^{n} \bar{s}\left(\frac{k-n\kappa+\gamma\sum_{k=1}^{n}t_k}{1+n\gamma} - t_i\right)$ & if $k_n^- \leq  k < k_n^+$ \\
	$\prod_{i=1}^{n} \bar{s}\left(\frac{k_n^+-n\kappa+\gamma\sum_{k=1}^{n}t_k}{1+n\gamma} - t_i\right)$ & if $k_n^+ \leq  k < k_{n+1}^-$ 
	\end{tabular}
	\right.
	\end{equation}
\end{widetext}
with $k_n^+ = n\kappa + (1+n\gamma)t_{n+1} - \gamma\sum_{k=1}^{n}t_k$ and  $k_n^-=k_{n-1}^+   + \kappa$. $\mathcal{F}_K(k)$ consists in an alternation of monotonically decreasing parts and of constant plateaus corresponding to the launch of a new searcher, as shown in figure \ref{fig:quantiles_N3}.

Let $\Omega$ be the $(N-1)$-dimensional space of possible values taken by $\left\{t_2, \cdots, t_N \right\}$. Because $\mathcal{F}_K(k)$ is a piecewise function of $k$ whose intervals are functions of the times $\left\{t_i \right\}i$, the solution $k_z$ of $\mathcal{F}_K(k_z) = z$ is also a piecewise function in $\Omega$. We note $\Omega_n$ the subspace of $\Omega$ such that $ k_z \in [k_n^- , k_n^+]$. Subdomains $\Omega_n$ and $\Omega_{n+1}$ are adjacent, i.e. they share a border $\Lambda_n$ defined as $k_z = k_n^+$, or equivalently $k_z = k_{n+1}^-$. A visual reprentation of these subdomains is shown in figure \ref{fig:quantiles_N3} (right panel).\\

In $\Omega_n$, $k_z$ is found as the solution of 
\begin{equation}
\prod_{i=1}^{n} \bar{s}\left(\frac{k_z-n\kappa+\gamma\sum_{k=1}^{n}t_k}{1+n\gamma} - t_i\right)  = z
\end{equation}
If there is a local minimum of $k_z$ in $\Omega_n$ it must be such that $\partial_{t_i} k_z = 0$ for all $i\leq n$. Differentiating this equation with respect to $t_i$ we find after some algebra 
\begin{equation}
(\partial_{t_i} k_z + \gamma)  \sum_{j=1 }^n \xi_j^{(n)}  -	(1+n\gamma)\xi_i^{(n)}  = 0
\end{equation}
where we have defined $\xi_i^{(n)} = \zeta\left(\frac{k_z-n\kappa+\gamma\sum_{k=1}^{n}t_k}{1+n\gamma} - t_i\right)$. Noting $\left\langle \xi^{(n)} \right\rangle = \frac{1}{n}\sum_{j=1}^n \xi_j^{(n)}$ the mean of all $\xi_j^{(n)}$, the condition  $\partial_{t_i} k_z = 0$ for all $i\leq n$ for the existence of a local minimum in $\Omega_n$ imposes
\begin{equation}
\label{eq:xi_i_n}
\xi_i^{(n)} = \left\langle \xi^{(n)} \right\rangle  \frac{n\gamma}{1 + n\gamma } \geq   \left\langle \xi^{(n)} \right\rangle 
\end{equation}
This can not be realized for all $i$ as all $\xi_i^{(n)}$ can not simultaneously be larger than their mean, which implies that there can not be a local minimum within $\Omega_n$. $k_z$ is therefore minimized at the boundaries of $\Omega_n$, i.e. one of the lower-dimensional subspaces $\Lambda_n$. There, the function $k_z(\left\{t_i\right\})$ is discontinuous, jumping from a value $k_n^+$ to a larger value $k_n^-$. The global minimum will therefore such that $k_z = k_n^+$ for a certain value of $n$.

Using the definition of $k_n^+ = n\kappa + (1+n\gamma)t_{n+1} - \gamma\sum_{k=1}^n t_k$ and substituting it into equation (\ref{eq:def_quantiles}), the equation that defines the $(N-2)$-dimensional subspace $\Lambda_n$ reads
\begin{equation}
\label{eq:s_tauin}
\prod_{i=1}^{n} \bar{s}\left(\tau_i^{(n)}\right)  = z
\end{equation}
where we have defined $\tau_i^{(n)}=t_{n+1}-t_i$. This constraints the value of one of the $\tau_i^{(n)}$ and we can choose in particular to express $\tau^{(n)}_1$ in terms of the other $\tau_i^{(n)}$, i.e.
\begin{equation}
\tau_i^{(1)} = \bar{s}^{-1}\left(\frac{z}{\prod_{i=2}^{n} \bar{s}\left(\tau_i^{(n)}\right)}\right)  
\end{equation}
Using this expression as well as $t_i = t_{n+1}-\tau_i^{(n)}$ we can rewrite $k_n^+$ as 
\begin{equation}
k_n^+ = n\kappa + (1+\gamma)  \bar{s}^{-1}\left( \frac{z}{\prod_{i=2}^{n} \bar{s}\left(\tau_{i}^{(n)} \right)}\right)  + \gamma \sum_{i=2}^{n}  \tau_i^{(n)}
\end{equation}
If there exists a local minimum in $\Lambda_n$, then it should be such that $\partial_{\tau_i^{(n)}} k_n^+ = 0$. After some algebra, we can write $\partial_{\tau_i^{(n)}} k_n^+$ as
\begin{equation}
\label{eq:diff_kn+}
\partial_{\tau_i^{(n)}} k_n^+ = \gamma - (1+\gamma)  \frac{\zeta\left(\tau_{i}^{(n)}\right)}{\zeta\left(\tau_{1}^{(n)} \right)}
\end{equation}
Note that for any $i>1$ we have $\tau_{1}^{(n)}>\tau_{i}^{(n)}$. If $\bar{s}$ is logarithmically convex, then $\zeta$ is a negative, monotonically growing function, i.e. $\zeta\left(\tau_{i}^{(n)}\right)<\zeta\left(\tau_{1}^{(n)} \right)\leq 0$, leading to $\partial_{\tau_i^{(n)}} k_n^+ < -1$. We therefore conclude that $k_n^+$ decreases monotonically with each $\tau_i^{(n)}$ such that there can not be any local extremum of $k_z$ within $\Lambda_n$. The minimum of $k_z$ has thus to be found for $\tau_i^{(n)} = t_{n+1}$, that is for $t_i=0$. \\

To find the global minimum of $k_z$ we finally need to compare $k_n^+$ for all values of $n$ where we set $t_i=0$ for $i\leq n$ and $t_{n+1} = \bar{s}^{-1}\left( z^{\frac{1}{n}} \right)$ as imposed by equation (\ref{eq:s_tauin}). This yields
\begin{equation}
\label{eq:kn+_opt}
k_n^+ = n\kappa + (1+n\gamma)\bar{s}^{-1}\left( z^{\frac{1}{n}} \right)
\end{equation}
We finally conclude that the optimal strategy that minimizes the $q$-quantile with $q=1-z^{-1}$ is therefore to initially launch $n_z^*$ searcher where $n$ minimizes the right-hand side of equation (\ref{eq:kn+_opt}) and launch the next searchers at times later than 
\begin{equation}
t_z^* = \bar{s}^{-1}\left( z^{\frac{1}{n_z^*}} \right)
\end{equation}   
This general result holds for any log-convex SASP.

\subsection{Subsequent quantiles}

Following the optimal strategy that minimizes $k_z$ as decribed in the previous paragraph, we note that the launch times $t_i$ for $i>n_z^*$ do not impact the optimized quantile as long as they are larger than $t_z^*$. One is therefore free to launch subsequent searchers arbitrarily without altering the result presented in the previous section. One might in particular want to optimize their launch times in order to minimize a subsequent $q'$-quantile with $z'=1-{q'}^{-1}<1-q^{-1} = z$, provided that the $q$-quantile has already been optimized. 

We can perform the same analysis as in the previous section by setting all $t_i=0$ for $i\leq n_z^*$ and check whether a local minimum with respect to later launch times can be found within one of the subdomains $\Omega_n$ with $n\geq n^*_z$. Following similar calculations, we show that the condition $\partial_{t_i} k_{z'} = 0$ within a subdomain $\Omega_n$ with $i\geq n \geq n_z^*$ leads to 
\begin{equation}
\label{eq:nextquantile_xi}
\xi_i^{(n)} = \left[\left\langle \xi^{(n)} \right\rangle_{n_z^*}^n + \frac{n_z^*}{n}\left( \xi_z^* - \left\langle \xi^{(n)} \right\rangle_{n_z^*}^n \right)\right] \frac{n\gamma}{1+n\gamma} 
\end{equation}
where we have defined $\left\langle \xi^{(n)} \right\rangle_{n_z^*}^n = \frac{1}{n-n_z^*}\sum_{j=n_z^*+1}^n \xi_j^{(n)}$ as the mean of the $\xi_j^{(n)}$ comprised between  $n_z^*+1$ and $n$, and $\xi_z^* = \xi_j^{(n)}$ for $j\leq n_z^*$. Using again the log-convexity of $\bar{s}$, we have $\xi_z^* > \left\langle \xi^{(n)} \right\rangle_{n_z^*}^n$ and in turn $\xi_i^{(n)} > \left\langle \xi^{(n)} \right\rangle_{n_z^*}^n$.  Using this inequality in (\ref{eq:nextquantile_xi}) we obtain 
\begin{equation}
\xi_i^{(n)} \geq \left\langle \xi^{(n)} \right\rangle_{n_z^*}^n 
\end{equation}
Similarly to equation (\ref{eq:xi_i_n}), this can not be realized for all $i>n_z^*$ and we conclude that there can not be a local minimum of $k_{z'}$  within one of the subdomains $\Omega_n$. The minimum must therefore be found in one of the boundaries $\Lambda_n$.  
In this case, following the same line of calculations that led to equation (\ref{eq:diff_kn+}), we obtain
\begin{equation}
\label{eq:diff_kn+_qprime}
\partial_{\tau_i^{(n)}} k_n^+ = \gamma - \frac{1+\gamma}{n_z^*}  \frac{\zeta\left(\tau_{i}^{(n)}\right)}{\zeta\left(\tau_{1}^{(n)} \right)}
\end{equation}
Here, we can not conclude in general on the sign of the latter quantity. Using the log-convexity of $\bar{s}$ we can simply bound $\partial_{\tau_i^{(n)}} k_n^+$ from above by $\gamma\left(1-{n_z^*}^{-1}\right) - 1$, which can take positive values if $\gamma$ is large enough.\\

However if $\gamma < \left(1-{n_z^*}^{-1}\right)^{-1}$, $k_{z'}$ can not be minimized within any of the $\Lambda_n$ and its minimum is found for $t_i = t_z^*$ for all $n_z^* < i\leq n$. 
In this case, the optimal strategy is therefore as follows. After having launched $n_z^*$ searchers at $t=0$ to minimize $k_z$, $n_{z,z'}^{*}$ new searchers should be launched at time $t_{z}^*$ to minimize $k_{z'}$, where $n_{z,z'}^{*}$ is defined as 
\begin{equation}
  n_{z,z'}^{*} = \text{argmin}_n\left[n\kappa  + (1+n\gamma) t^{(n)}_{z,z'} \right]- n_z^*
\end{equation} 
and $t^{(n)}_{z,z'}$ is defined through the equation (\ref{eq:s_tauin}) that in this situation reads
\begin{equation}
\label{eq:Lambdan_zprime}
\bar{s}\left(t^{(n)}_{z,z'} - t_z^* \right)^{n-n_z^*}\bar{s}\left(t^{(n)}_{z,z'}\right)^{n_z^*} = z'
\end{equation}
Later searchers should then be introduced at a time later than $t_{z,z'}^{*}= t^{(n_{z,z'}^{*}+n_z^*)}_{z,z'}$. The procedure can then be repeated to again minimize subsequent quantiles.

\subsection{Canonical example: the exponential SASP}

Let us illustrate the results of the previous sections on a specific example, namely for an exponentially decaying SASP $\bar{s}(t) = e^{-\lambda t}$. Exponential decays are in fact widely observed in actual search processes, especially in confined space. The short-time behavior may differ (see \ref{section:numerical_optimization} for detailed discussion), but we choose here a simple functional form that allows analytic calculations for the sake of illustration. Similar calculations can still be performed for any log-convex SASP.

Here, the boundaries $\Lambda_n$ defined through equation (\ref{eq:s_tauin}) are found as 
\begin{equation}
t_{n+1} = \frac{1}{n}\left[ \sum_{k=1}^n t_k - \lambda^{-1}\ln(z) \right]
\end{equation}
We show them as lines in the $(t_2,t_3)$-plane for $N=3$ in figure \ref{fig:quantiles_N3}. There, the minimum of $k_z$ is found either in the $\Omega_1$-subdomain where $k_z$ is constant, or on the line $t_2=0$ in $\Omega_2$ as $k_z$ does not depend on $t_3$ there, or at $t_2=t_3=0$. In either of these locations, we have $k_z=k_n^+$ with 
\begin{equation}
k_n^+ = n\kappa - \left(\frac{1}{n} + \gamma\right)\ln(z)
\end{equation}
The value $n_z^*$ that minimizes this quantity and the corresponding launch time $t_z^*$ are found as 
\begin{equation}
n_z^* = \sqrt{\frac{-\ln (z)}{\lambda \kappa}}, \ \ t_z^* = \sqrt{\frac{-\kappa \ln(z)}{\lambda}}
\end{equation}
Optimally, $n_z^*$ searchers should be introduced at $t=0$ while the next searchers should be introduced at a time later than $t_z^*$. Interestingly, this result does not depend at all on $\gamma$. Whatever the cost for sustaining the searchers, it will not impact how many should be launched initially in the optimal strategy. \\

When it comes to launching later searchers in order to minimize a subsequent quantile $k_{z'}$, we first need to evaluate equation (\ref{eq:diff_kn+_qprime}), reading $\partial_{\tau_i^{(n)}} k_n^+ = \gamma - (1+\gamma)/n_z^*$ for $i\leq n < n_z^*$. If $\gamma >{n_z^*}/(n_z^*-1)$, then $\partial_{\tau_i^{(n)}} k_n^+>0$ and the minimum is found for $\tau_i^{(n)} = 0$, i.e. $t_i = t_{n+1}$. In this case, $k_n^+$ increases linearly with $n$ such that the value $n\geq n_z^*$ that minimizes it is $n=n_z^*$: optimally no new searcher should be introduced. On the other hand, if $\gamma <{n_z^*}/(n_z^*-1)$ the minimum of $k_z$ in each $\Lambda_n$ is found for $t_i= t_z^*$. In this case, equation (\ref{eq:Lambdan_zprime}) reads $t^{(n)}_{z,z'} = t_z^*\left(1 - n_z^*/n \right)-\ln(z')/n\lambda$ and we can show that $k_n^+$ is minimized for a value  
\begin{equation}
m_{z,z'}^* =  \sqrt{\frac{-\ln(z) - \ln(z')}{\kappa\lambda + \gamma \sqrt{-\kappa\lambda \ln(z)}  }}
\end{equation}
If this number is larger than $n_z^*$ one should launch $n_{z,z'}^*= m_{z,z'}^*-n_z^*$ new searchers at time $t_{z,z'}^{(m_{z,z'}^*)}$. Otherwise, we launch no new searcher at time $t_{z,z'}^{(n_z^*+1)}$. Again, we can then continue to look for optimal ways to launch the next searchers and minimize subsequent quantiles. \\

\section{Optimizing the mean search cost}

While the quantiles of the search cost quantify the probability of not reaching too large search costs, they do not account much for the long-time behavior of the SASP. In contrast, the mean search cost does account for it and is also highly relevant in many processes.  Given the joint first-passage time distribution $\bar{R}(t) = -\bar{S}'(t)$, the mean first-passage time $\bar{T}$ is classically obtained as $\bar{T} = \int_0^\infty t \bar{R}(t) dt = \int_0^\infty \bar{S}(t) dt$ using integration by parts. Similarly, we have 
\begin{align}
\bar{\mathcal{T}} =&   \sum_{n=1}^\infty \int_{0}^\infty \frac{d}{dt}\left((t-t_n)\Theta(t-t_n)\right) \bar{S}(t) dt  \nonumber \\
=&  \sum_{n=1}^\infty n  \int_{t_n}^{t_{n+1}} \bar{S}(t) dt
\end{align}
and finally 
\begin{align}
\bar{\mathcal{N}} =&   \sum_{n=1}^\infty \int_{0}^\infty \frac{d  \Theta(t-t_n)}{dt} \bar{S}(t) dt = \sum_{n=1}^\infty \bar{S}(t_n) 
\end{align} 
Combining the contributions, we obtain the mean search cost as 
\begin{equation}
\bar{K}(\{ t_i\}) = \sum_{n=1}^{N} \left[\kappa \bar{S}\left(t_n\right) + (1+n\gamma) \int_{t_{n}}^{t_{n+1}} \bar{S}\left(t\right)  dt \right]\;, 
\end{equation}
Minimizing this quantity in general for an arbitrary SASP is not possible. However, we can still derive interesting results if we assume that there exists only one local minimum of $\bar{K}$. In this section, we will first take a detour through the case $N=2$ in order to get an intuition for the conditions for such a unique minimum to exist. We will then derive general results under this assumption, and finally perform numerical optimization for certain test cases.

\subsection{Two searchers}

For two searchers, the mean search cost is a function of only one variable, namely the launch time $t_2$ of the second searcher. Although we do not aim at determining analytically the optimal value for $t_2$ in general, we want to understand the conditions for which a unique optimal strategy can be found.

The mean search cost $\bar{K}$ of the 2-searcher process reads as follows:
\begin{align}
\bar{K} =& \kappa(1+\bar{s}(t_2)) + (1+\gamma) \int_{0}^{t_2} \bar{s}\left(t\right)dt \nonumber \\
&+ (1+2\gamma)\int_{0}^\infty  \bar{s}\left(t\right)\bar{s}\left(t+t_2\right) dt 
\end{align}
Upon integration by parts, the first derivative of $\bar{K}$ can be written as 
\begin{align}
\label{eq:dK2_dt2}
\bar{K}'(t_2) =& \kappa \bar{s}'(t_2) -  \gamma \bar{s}\left(t_2\right)  - (1+2\gamma) \int_{0}^\infty  \bar{s}'\left(t\right)\bar{s}\left(t+t_2\right) dt 
\end{align}
The extrema of $\bar{K}$ are found such that $\bar{K}' = 0$. For reasons that will become apparent later, we rewrite this condition as
\begin{align}
\label{eq:kappa_dK2}
\kappa  = \frac{ \gamma \bar{s}(t_2) + (1+2\gamma) \int_{0}^\infty  \bar{s}'\left(t\right)\bar{s}\left(t+t_2\right) dt}{\bar{s}'(t_2)}
\end{align}
Differentiating equation (\ref{eq:dK2_dt2}) with respect to $t_2$  and substituting $\kappa$ with the right-hand side of equation (\ref{eq:kappa_dK2}) we can write $\bar{K}''(t_2)$ as 
\begin{align}
\bar{K}''(t_2) = -\bar{s}'(t_2) \frac{d}{dt_2} \left[  \frac{g(t_2)}{\bar{s}'(t_2)} +\gamma \frac{\bar{s}(t_2) + g(t_2)}{\bar{s}'(t_2)} \right]
\end{align} 
where we have defined
\begin{align}
g(t_2) &= \int_{0}^\infty  \bar{s}'\left(t\right)\bar{s}\left(t+t_2\right) dt 
\end{align}
First, we have 
\begin{align}
\label{eq:dh1_dt2}
\frac{d}{dt_2}  \left(\frac{g}{\bar{s}'}\right) =& \int_{0}^{\infty} \bar{s}'(t)\frac{\bar{s}'(t+t_2)\bar{s}'(t_2)-\bar{s}''(t_2)s(t+t_2)}{\bar{s}'(t_2)^2} dt
\end{align}
If $\bar{s}$ is log-convex, then for any $x,y \in \mathbb{R}^+$ with $x<y$ it holds $\bar{s}''(x)s(y)>\bar{s}'(x)\bar{s}'(y)$ such that the right-hand side of equation (\ref{eq:dh1_dt2}) is positive.

Then, we can write upon integration by parts  
\begin{align}
\frac{\bar{s}(t_2) + g(t_2)}{\bar{s}'(t_2)}  =&  2  \int_{0}^\infty  \frac{s(t+t_2) -\bar{s}\left(t\right)\bar{s}(t_2)}{\bar{s}(t_2)} \frac{\bar{s}'\left(t+t_2\right)}{\bar{s}'(t_2)} dt  
\end{align}
Now, note that 
\begin{align}
\frac{d}{dt_2} &\frac{s(t+t_2) -\bar{s}\left(t\right)\bar{s}(t_2)}{\bar{s}(t_2)} \nonumber \\
&\ \ \ \ \ \ = \frac{\bar{s}'(t+t_2)\bar{s}(t_2) -\bar{s}\left(t+t_2\right)\bar{s}'(t_2)}{\bar{s}(t_2)^2}
\end{align}
and 
\begin{equation}
\frac{d}{dt_2} \frac{\bar{s}'\left(t+t_2\right)}{\bar{s}'(t_2)} = \frac{\bar{s}''(t+t_2)\bar{s}'(t_2)-\bar{s}'(t+t_2)\bar{s}''(t_2)}{\bar{s}'(t_2)^2}
\end{equation}
Assuming log-convexity for both $\bar{s}$ and $-\bar{s}'$, both these terms are positive which implies that the integrand in $h$ is the product of two positive increasing functions of $t_2$, which makes $h$ also a positive, increasing function of $t_2$. This proves that the derivative of $(\bar{s} + g)/\bar{s}'$ is positive.\\

We therefore conclude that $\bar{K}''$ is the sum of two positive terms, i.e. $\bar{K}''>0$ for any local extremum of $\bar{K}$. Any local extremum is therefore a local minimum. Now, suppose that there are two local minima. By continuity and differentiability, there must also be a local maximum between both of them, which is impossible. We conclude that $\bar{K}$ has at most one local minimum and no local maximum. However, we cannot determine the value of $t_2$ that minimizes $\bar{K}$ in general as equation (\ref{eq:dK2_dt2}) can not be inverted for an arbitrary SASP $\bar{s}(t)$.\\

Unlike the case of quantiles, where the log-convexity of the survival probability allowed identifying optimal strategies for an arbitrary number of searchers $N$, it is much more difficult to prove that this assumption is sufficient for the uniqueness of a local minimum of the mean search cost for arbitrary values of $N$. However, in the next sections we will show numerically that this condition still holds for $N>2$.

\subsection{Conditions for simultaneous launching}

Although we cannot analytically predict the details of optimal launch procedures for arbitrary SASPs, we can derive a criterion to decide whether it is beneficial to launch searchers simultaneously or not in order to minimize the mean search cost. Before we proceed, let us first introduce some notations:
\begin{align}
s_{k} =& \bar{S}(t_k) \\
I_{nk} =& \int_{t_n}^{t_{n+1}} \bar{S}(t) \zeta(t-t_k) dt \\
r_{nk} =& s_k \zeta(t_n-t_k)
\end{align}
With these notations the gradient of $\bar{K}$ reads 
\begin{align}
\label{eq:grad_K}
\nabla_{k}\bar{K} = -\gamma s_{k} &- \sum_{n=k}^{N} (1+n\gamma) I_{nk} \nonumber \\  
&+ \kappa \left( \sum_{n=1}^{k-1} r_{kn} - \sum_{n=k+1}^{N} r_{nk} \right)
\end{align}
where $\nabla_{k} = \partial_{t_k}$. We also note $\mathbf{t}_N = \left\{t_2, \cdots, t_N \right\}$.\\

Let us first assume that the mean search cost $\bar{K}(\mathbf{t}_N)$ has at most one unique local minimum located at a point $\mathbf{t}_N^*$. At this location, two possible configurations can in principle be observed for the launch times $i$. Either $t_i^*$ is strictly larger than $t_{i-1}^*$ and the gradient $\nabla_i \bar{K}_N(\mathbf{t^*_N})$ vanishes, i.e. it is an actual local extremum of $\bar{K}$ with respect to $t_i$. Or $t_i^*=t_{i-1}^*$, i.e. the $i$\textsuperscript{th} and $(i-1)$\textsuperscript{th} searchers are launched simultaneously, which happens only if the derivative of $\bar{K}$ with respect to $t_i$ at this location is positive.  Let us now proceed by contradiction. 

Assume that $\bar{K}$ is minimized at a point $\mathbf{t}^*_N$ where $t_{k+1}^*=t_{k}^*$ and $t_{k}^*> t_{k-1}^*$ for a certain $k>2$. Following our previous observation, this implies that $\nabla_{k+1} \bar{K} > 0$ and $\nabla_{k} \bar{K} = 0$. In addition, for $t_{k+1}=t_{k}$ we have $s_{k} = s_{k+1}$, $I_{n,k} = I_{n, k+1}$, $r_{k,n} = r_{k+1,n}$, $r_{n,k} = r_{n,k+1}$ and $I_{k,k} = 0$. When computing the difference between $\nabla_{k} \bar{K}$ and $\nabla_{k+1} \bar{K}$ with these constraints, only two terms remain, namely
\begin{align}
\label{eq:diff_grad}
\nabla_{k} \bar{K}-\nabla_{k+1} \bar{K}=& -\kappa \left( r_{k+1,k} + r_{k+1,k} \right)\nonumber \\
=& - 2\kappa \bar{s}'(0) S_k(t_k) > 0
\end{align} 
Because $\bar{s}'(0)<0$, we therefore have $\nabla_{k}\bar{K}_N > \nabla_{k+1}\bar{K}_N$. This inequality is valid at any point $\mathbf{t}_N$, provided that $t_{k+1}=t_{k}$, and in particular at $\mathbf{t}^*_N$, where we have $\nabla_{k+1} \bar{K}_N > 0$. We thus obtain that $\nabla_{k}\bar{K} > \nabla_{k+1}\bar{K} >0$ at the minimum of $\bar{K}$, which is in contradiction with the original assumption. 

We therefore conclude that if $t_k^*=t_{k-1}^*$ then for all $j<k$ we must have $t_k^* = t_j^* = 0$: in the optimal strategy, if searchers are launched simultaneously it must necessarily be at the start of the process and not later.  \\

From this conclusion, we can now identify the optimal number of agents $N_{sim}$ that should be launched simultaneously at $t=0$. To do this, we note that the optimal strategy must be such that $\nabla_p \bar{K}_N \geq 0$ for $p \leq N_{sim}$ and $\nabla_p \bar{K}_N=0$ for $p>N_{sim}$. We will therefore calculate the derivative $\nabla_k \bar{K}$ for an arbitrary $k$ with two constraints, namely $t_p=t_k=0$ for $p\leq k$ and $\nabla_p \bar{K}_N=0$ for $p>k$. If this derivative is positive, we know from equation (\ref{eq:diff_grad}) that all previous derivatives for $p<k$ are also positive. We therefore identify $N_{sim}$ as the largest value of $k$ for which $\nabla_k \bar{K}$ is positive under the mentioned constraints.  

We first evaluate $\nabla_k \bar{K}$ with $t_2=\cdots=t_k=0$ using equation (\ref{eq:grad_K}) and note that some terms can be simplified, namely
\begin{align}
\label{eq:sk_1}
&s_k = 1\\
\label{eq:Ink_simp}
&I_{nk} = \int_{t_n}^{t_{n+1}} \bar{s}'(t) \bar{s}(t)^{k-1} \prod_{p=k+1}^n s(t-t_p)  dt \text{ for } n>k \\
\label{eq:rnk_simp}
&\sum_{n=1}^{k-1} r_{kn} = (k-1)\bar{s}'(0)  \text{ for } n>k 
\end{align} 
The integral term $I_{nk}$ can be transformed using integration by parts, yielding
\begin{align}
I_{nk}   =& \frac{1}{k} \left[s_{n+1} - s_n \right] - \frac{1}{k} \sum_{p=k+1}^n I_{np}
\end{align}
Now, summing over all values of $n\geq k$ yields
\begin{align}
\sum_{n=k}^{N}  (1+n\gamma) I_{nk} =& - \frac{1}{k}  - \gamma -\frac{\gamma}{k} \sum_{n=k+1}^{N}  s_n  \nonumber \\
&-  \frac{1}{k} \sum_{p=k+1}^{N} \sum_{n=p}^{N} (1 + n\gamma)   I_{np}
\label{eq:intermediate_condsim}
\end{align}
where we have used $s_k=1$ and $s_{N+1} = 0$. We now use the condition $\nabla_p\bar{K} = 0$ for $p>k$, which can be written using equation (\ref{eq:grad_K}) as 
\begin{align}
\gamma s_p  &+ \sum_{n=p}^{N}  (1+n\gamma)  I_{np}  - \kappa \left(\sum_{n=1}^{p-1} r_{pn}  - \sum_{n=p+1}^{N} r_{np} \right)=0
\label{eq:cond_gradp_0}
\end{align}
Now, we note that the summand in the last term of equation (\ref{eq:intermediate_condsim}) is exactly the second term in the gradient of $\nabla_p \bar{K}$ in equation (\ref{eq:cond_gradp_0}). Summing equation (\ref{eq:cond_gradp_0}) over $p>k$ yields 
\begin{align}
\sum_{p=k+1}^{N} \sum_{n=p}^{N} (1 + n\gamma)  I_{np}  =& - \sum_{p=k+1}^{N} \gamma s_{p} \nonumber \\
+ \kappa \sum_{p=k+1}^{N} & \left(\sum_{n=1}^{p-1}  r_{pn}  - \sum_{n=p+1}^{N} r_{np} \right)
\end{align}
By swapping the order of summation in the terms proportional to $\kappa$ we can show 
\begin{align}
\sum_{p=k+1}^{N}  \left(\sum_{n=1}^{p-1}  r_{pn}  - \sum_{n=p+1}^{N} r_{np} \right) =& k \sum_{p=k+1}^{N}  r_{pk}
\end{align}
where we have used the fact that $r_{pn} = r_{pk}$ for $n\leq k$. We therefore obtain 
\begin{align}
\label{eq:sum_Inp}
\sum_{p=k+1}^{N} \sum_{n=p}^{N} (1 + n\gamma)  I_{np} =& - \sum_{p=k+1}^{N} \gamma s_{p}   +  k \kappa \sum_{p=k+1}^{N}  r_{pk} 
\end{align}
Now, substituting equation (\ref{eq:sum_Inp}) into equation (\ref{eq:intermediate_condsim}) and inserting equations (\ref{eq:sk_1}), (\ref{eq:rnk_simp}) and (\ref{eq:intermediate_condsim}) into (\ref{eq:grad_K}) finally yields
\begin{align}
\label{eq:gradK_simult}
\nabla_k \bar{K}=&  \frac{1}{k} + \kappa (k-1)  \bar{s}'(0)  
\end{align}
$N_{sim}$ is therefore found as the largest value of $k$ for which this quantity is positive, i.e. 
\begin{equation}
\label{eq:Nsim}
N_{sim} = \left\lfloor \frac{1}{2}+\sqrt{\frac{1}{2}-\frac{1}{\kappa \bar{s}'(0) }} \right\rfloor \stackrel{|\kappa \bar{s}'(0)|\to 0}{\simeq} \frac{1}{\sqrt{-\kappa \bar{s}'(0)}}
\end{equation}
Surprisingly, $N_{sim}$ does not depend on $\gamma$ at all: no matter how much it costs to sustain a searcher, the number of agents to be introduced into the system at $t=0$ will only be governed by the launch cost $\kappa$ and the initial slope of $s'$, i.e. the initial value of the first-passage time distribution: if $\bar{s}(t)$ is very sharply decreasing at short times --  i.e. if the probability to find the target quickly is high -- there is no interest in launching multiple searchers initially as the benefit in the search time would be overcompensated by the launch cost.

\subsection{Numerical optimization}
\label{section:numerical_optimization}

\subsubsection{Test cases}

While the statistical properties of the search cost depend on the details of the SASP, in a vast majority of single-agent search processes $\bar{s}(t)$ has a functional form that falls into only a few different classes. At long times $\bar{s}(t)$ either decays exponentially (e.g. in confined domains \cite{mattos2012first, shaebani2018unraveling}) or algebraically (e.g. in open space \cite{bray2013persistence}). Faster decays are extremely rare. Note that we consider here only SASP whose probability to eventually find the target is 1, i.e. $\lim_{t\to\infty}\bar{s}(t)=0$. At short times, three cases can be distinguished. First, the initial slope of $s'(t)$ can be very low and asymptotically close to $0$. This happens for searches that can not be infinitely fast, e.g. when the target and the searchers are initially always at a finite distance from one another, and leads to non-monotonic -- and non-convex -- first-passage time distributions \cite{godec2016first, redner2023first}. Second, $-s'(0)$ is finite but sufficiently large for $s(t)$ to be convex. This is a frequent feature of search processes with random initial positions of the searchers \cite{koren2007leapover}. Third, $s'(t)$ decays infinitely fast at $t=0$ and the first-passage time distribution diverges at $t=0$, as it is observed in some simple diffusive processes \cite{slepian1962one}.  \\

\begin{figure}
	\centering
	\includegraphics[width=\linewidth]{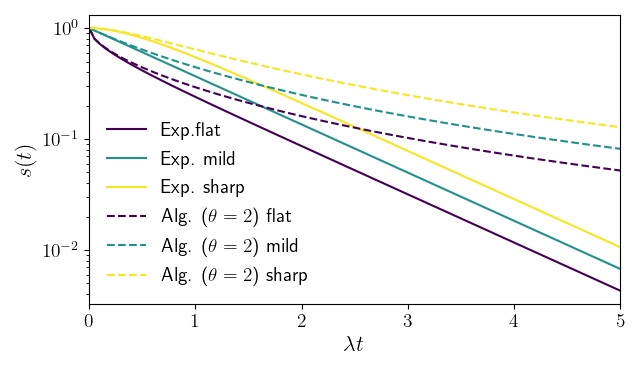}
	\includegraphics[width=\linewidth]{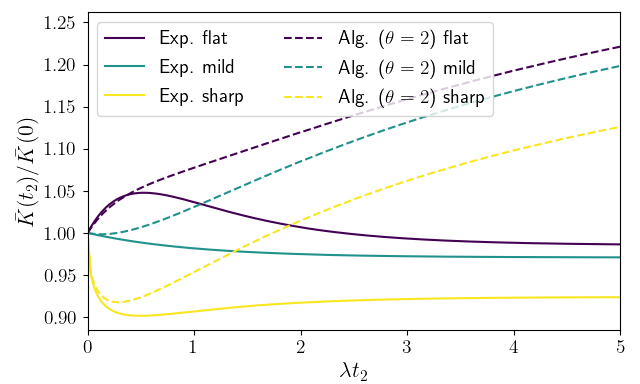}
	\caption{$\underline{\text{Top}}$: Test cases of single-agent survival probabilities for the numerical optimization of the mean search cost. They combine two long-time with three short-time behaviors are allow to discriminate between possible optimal launch strategies. $\underline{\text{Bottom}}$: Mean search cost $\bar{K}$ as a function of $t_2$, normalized by its value for $t_2=0$, for the six test cases of SASP. We use here $\kappa\lambda=0.55$ and $\gamma=0.1$. Some SASP lead to non-trivial minima for the mean cost while others are minimized for $t_2=0$ pr $t_2\to\infty$.}
	\label{fig:test_SASP}
\end{figure}

In order to reach a broad understanding of possible optimal launch strategies, we construct test cases which combine specific short- and long-time behaviors. To do this, we compose $t \mapsto e^{-\lambda t}$ and $t \mapsto (1+\lambda\theta^{-1} t)^{-\theta}$ characterizing the long-time behavior with either $x\mapsto \sin\left(\pi x/2 \right)$, $x\mapsto 1$ or $x\mapsto 2\arcsin\left(x\right)/\pi$ that characterize the short-time behaviors. The six resulting SASP we focus on are therefore 
\begin{align}
s_{exp}^{mild}(t) =& e^{-\lambda t} \\
s_{exp}^{flat}(t) =& \sin\left(\frac{\pi}{2} e^{-\lambda t} \right) \\
s_{exp}^{sharp}(t) =& \frac{2}{\pi} \arcsin\left( e^{-\lambda t} \right) \\
s_{alg}^{mild}(t) =& \frac{1}{\left(1 + \lambda t\theta^{-1} \right)^\theta} \\
s_{alg}^{flat}(t) =& \sin\left( \frac{\pi}{2\left(1 + \lambda t\theta^{-1} \right)^\theta} \right) \\
s_{alg}^{sharp}(t) =& \frac{2}{\pi} \arcsin\left( \frac{1}{\left(1 + \lambda t\theta^{-1} \right)^\theta} \right) 
\end{align}
In practice we used $\theta=2$. We show these six function in figure \ref{fig:test_SASP} for visualization, as well as the mean search cost $\bar{K}$ as a function of $t_2$ in a 2-searcher process. \\

To identify optimal launch strategies, we performed standard gradient descent optimization. We start with $N=2$ and we initialize $t_2$ to a random value $t_2 = t_2^{(0)}$. At each step, we calculate the gradient of $\bar{K}$ with respect to $t_2$ and update $t_2^{(n+1)} = t_2^{(n)} - r \partial_{t_2} \bar{K}|_{t_2 = t_2^{(n)}}$ where $r$ is the learning rate. We stop the iteration either if the $t_2$ has converged to a constant value or if $t_2=0$ and $\partial_{t_2} \bar{K}>0$. We then add a new degree a freedom $t_3$ that we initialize to a random value $t_3^{(0)}>t_2$. At each step we now compute the full gradient of $\bar{K}$ with respect to $t_2$ and $t_3$ which we use to update both of the degrees of freedom until convergence. We continue this process by adding more and more degrees of freedom until we reach $N-1$ degrees of freedom that are such that $\bar{S}(t_N)<0.001$. 

Among the six test cases under study, four of them had unique local minima of the mean search cost, namely the {\it mild} and {\it sharp} cases.  They correspond to log-convex SASP, which is consistent with the prediction for 2 searchers. No local maximum was found such that equation (\ref{eq:Nsim}) holds for these cases. The two other cases, i.e. non-convex SASPs, led to more complex structures for the mean search cost and are analyzed separately.

\begin{figure*}
	\begin{center}
		\includegraphics[width=.325\linewidth]{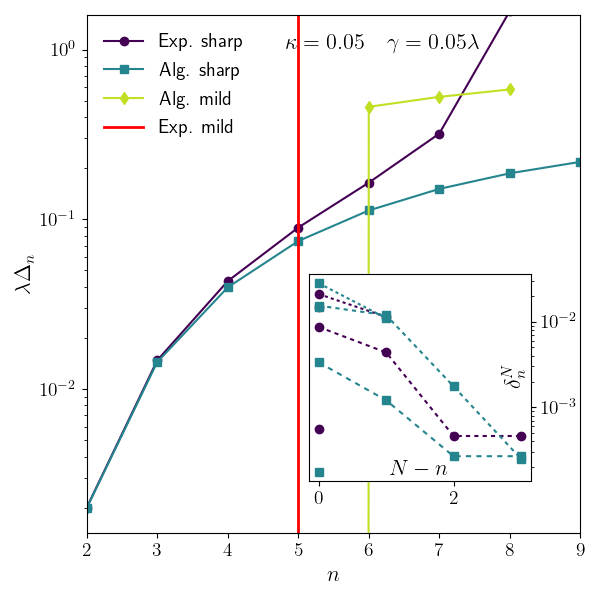}
		\includegraphics[width=.325\linewidth]{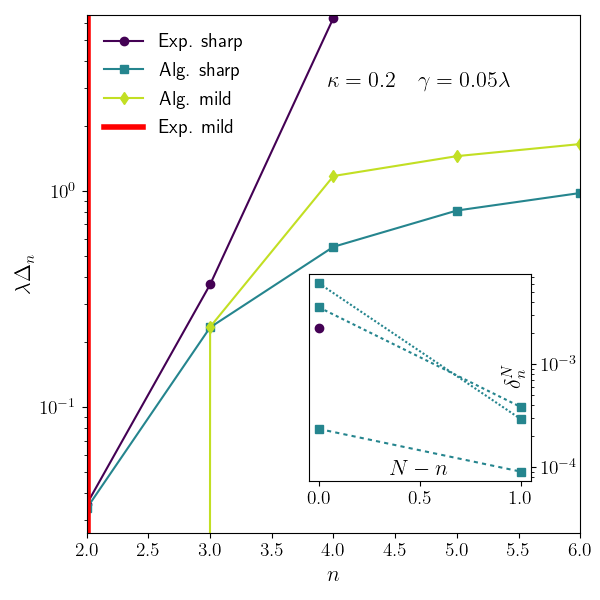}
		\includegraphics[width=.325\linewidth]{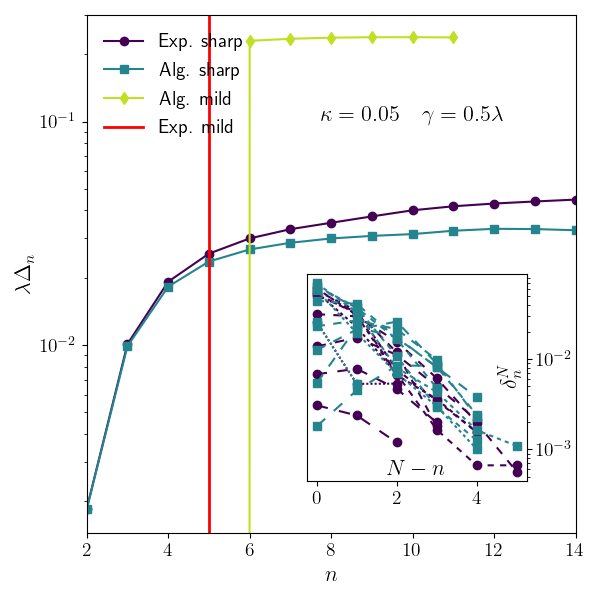}
		\includegraphics[width=.495\linewidth]{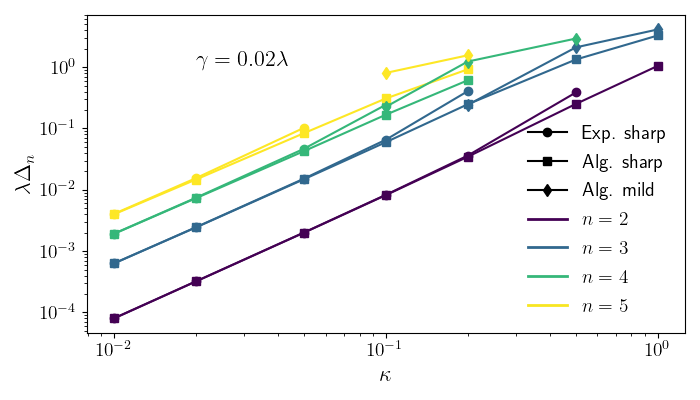}
		\includegraphics[width=.495\linewidth]{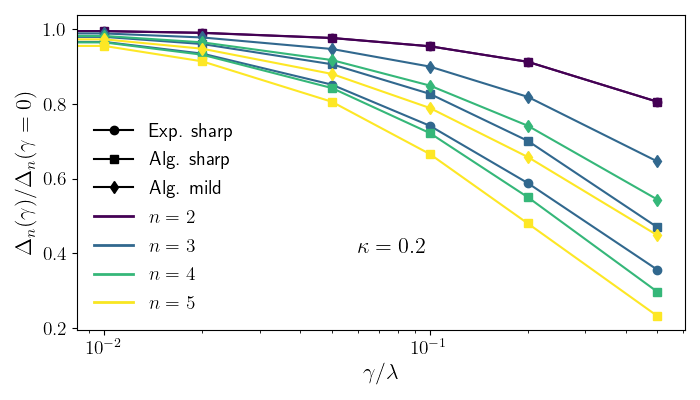}
		\caption{$\underline{\text{Top}}$: Launch intervals $\Delta_{n}$ normalized by $\lambda$ in the optimal strategies, as identified by numerical optimization, as a function of the index of the $n$\textsuperscript{th} agent. Different panels correspond to different values of $\kappa$ and $\gamma$ while different curves correspond to different SASP. The inset shows the relative variation $\delta_n^N$ as a function of $N-n$, for which $\Delta_n^{N+1}-\Delta_n^{N+1}$ was found larger than the tolerance of the gradient descent algorithm. We find that the optimal launch interval for low $n$ is governed by the short-time behaviour of $s(t)$. For mild ones, multiple searchers may be launched initially, which is shown here with vertical lines as $\Delta_n = 0$ for $n\leq N_{sim}$. For sharp ones, only one searcher is launched at $t=0$ and subsequent searchers are launched at later times. For large $n$, the launch interval is governed by the long-time behaviour of $s(t)$. For algebraic ones, $\Delta_n$ reaches a constant value while for exponential ones, $\Delta_n$ diverges. We do not observe this divergence on the last plot as it occurs at very large values of $n$.	$\underline{\text{Bottom}}$: Optimal introduction intervals $\Delta_{n}$ as a function of $\kappa$ (left) and $\gamma$ (right) for three test cases. The values are normalized by $\Delta_n$ for $\gamma=0$ in the second plot for better visualization. We find that optimal launch intervals increase with $\kappa$ but decrease with $\gamma$. }
		\label{fig:numerical_optimization}	
	\end{center}
\end{figure*}

\subsubsection{Convex SASP} 

We first start with the convex test cases. Before we proceed with the numerical optimization, we first treat the mild exponential case $\bar{s}(t) = e^{-\lambda t}$ separately as it can be fully analyzed analytically. In this case we have 
\begin{align}
\bar{T} =&    \lambda^{-1} \left[1 - \sum_{n=2}^\infty     \frac{e^{-\lambda  \sum_{k=1}^{n-1} t_{n} -   t_k }}{n(n-1)}  \right]\\
\bar{\mathcal{T}} =&   \lambda^{-1}\\
\bar{N} =&  1+\sum_{n=2}^\infty  e^{-\lambda\sum_{k=1}^{n} (t_n-t_k)}
\end{align}
Now, note that $\sum_{k=1}^{n} (t_n-t_k) = \sum_{l=1}^{n-1} l \Delta_l $, where $\Delta_k = t_{k+1} - t_{k}$ is the time interval between two consecutive launches. $\bar{K}$ can thus be fully expressed as a function of these time intervals and its partial derivative with respect to $\Delta_k$ reads then 
\begin{equation}
\frac{\partial \bar{K}}{\partial\Delta_k} = \lambda k \sum_{n=k+1}^{\infty} e^{-\lambda \sum_{l=1}^{n-1} l\Delta_l } \left(\frac{\lambda^{-1}}{n(n-1)} - \kappa \right) 
\end{equation}
Here, the terms for which $\kappa\lambda n(n-1)<1$ contributes positively to the gradient. Let $N_{sim} = \left\lfloor  \frac{1}{2}\left(1+\sqrt{1+\frac{4}{\kappa \lambda }}\right) \right\rfloor$ be the lowest value of $n$ such that the latter equality is verified. Then,  for any $k\geq N_{sim}$ the derivative $\partial_{\Delta_k} \bar{K}$ is negative such that the optimal strategy is such that $\Delta_k \to\infty$. In this limit, the contributions of all terms with $n\geq N_{sim}$ vanish because of the exponential factor. Therefore, for all $k<N_{sim}$ the gradient will be positive such that the cost will be minimized for $\Delta_k = 0$. The optimal strategy is therefore such that $N_{sim}$ searchers should be launched at $t=0$ and none later. In this case, the mean search cost is found as
\begin{align}
\bar{K}_{opt} =& \gamma + \sqrt{\kappa \lambda} \left(\frac{2}{\sqrt{\kappa \lambda}+\sqrt{\kappa \lambda+4}} +   \frac{\sqrt{\kappa \lambda}+\sqrt{\kappa \lambda+4}}{2} \right)
\end{align}
which is obtained by neglecting the floor part in $N_{sim}$. As $\kappa\to0$, the terms between brackets tend to 1 and the overall cost grows as $\sqrt{\kappa\lambda}$. However, as $\kappa\lambda \gg 1$, the brackets are dominated by the second term which tend to $\sqrt{\kappa\lambda}$, making the cost grow linearly with $\kappa \lambda$.\\

For the three other convex test cases, we perform the analysis numerically. First we investigate to which extent the total number of available agents impacts the optimal introduction times by defining the relative variation $\delta_{n}^N = 1-\Delta_{n}^{N+1}/\Delta_{n}^N$ upon having a new available searcher, where the superscript $N$ refers to the total number of searchers.  As shown in the insets of Fig. \ref{fig:numerical_optimization}, $\delta_n^N$ decays roughly exponentially with $N-n$ starting with a relatively low amplitude. The optimal launch times $t_n$ obtained in the limit $N\to\infty$ are therefore a good approximation of the actual optimal ones for a finite number $N$ of available searchers.

In the limit $N\to\infty$, our results allow to identify the optimal launch strategies and to classify them, as shown in the top panel of Fig. \ref{fig:numerical_optimization}. We observe that the short-time behavior governs the launch time of the first agents while the long-time behavior governs the introduction of later agents. For the initially {\it mild} SASP, we have $N_{sim}\geq 1$ in accordance with equation (\ref{eq:Nsim}), i.e. multiple walkers may optimally be introduced simultaneously at the start of the process. For the {\it sharp} SASP, it holds on the contrary $N_{sim}=1$: one necessarily has to wait a certain time before launching a second walker. For searchers introduced at later times, the optimal launch intervals $\Delta_n$ diverge for exponentially decaying SASP: as one adds more and more searchers in the system, the probability to find the target in a short amount of time becomes so high that it would cost more to introduce a new searcher quickly than to simply wait for the ones already lauched to find the target. On the contrary, $\Delta_n$ reaches a constant value as $n$ grows for algebraically decaying SASP. Here, even as one adds new searchers in the system, the probability to wait a long time may still remain non-negligible. It is therefore preferable to launch searchers at a constant pace even if it is at a certain cost.

The number of agents around which the transition between {\it early} and {\it late} agents occurs decreases with $\kappa$ and increases with $\gamma$ and roughly corresponds to introduction times $t_n$ such that $\lambda t_n \sim 1$. This is consistent with the dependence of the optimal launch intervals with the cost contributions, namely it increases with $\kappa$ and decreases with $\gamma$ as shown in he lower panel of figure \ref{fig:numerical_optimization}. While the dependence on $\kappa$ is not surprising as a large launch cost should reward to wait longer before launching a new searcher, the dependence on $\gamma$ is less intuitive. It is in fact preferable to launch searchers at a higher frequency when the {\it sustaining} rate $\gamma$ is larger, indicating that the gain in the overall search time overcompensates the larger rate of resources required to sustain the new searchers. This trend holds for all tested SASP and appears to be a general result for a wide variety of search processes.

\subsubsection{Non-convex case}

The case of non-convex SASP is more complex to treat as multiple local minima may exist. A first straightforward {\it locally} optimal strategy can be identified using equation (\ref{eq:gradK_simult}). This equation still holds for non-convex SASP but the conclusion that we draw from it is different. In fact, if $s'(0)>-\frac{1}{\kappa k (k-1)}$, the gradient $\nabla_k\bar{K}$ for $t_2 = \cdots = t_k = 0$ is negative, provided that the derivatives with respect to later launch times vanish. This is in particular true for any $k$ if $s'(0)=0$. There, for any $k\leq N$, there exists a {\it locally} optimal strategy that consists in launching $k$ searchers initially and the next searchers later, i.e. $\bar{K}$ is at a local minimum at this location. To rationalize this, we note that since the probability for the first searcher to find the target at very short times is low, there is no gain in waiting a short amount of time for launching next searchers compared to launching it together with the first one. However, this locally optimal strategy is not necessarily the globally best one, especially if the launch cost $\kappa$ is high. 

\begin{figure*}
	\begin{center}
		\includegraphics[width=0.49\linewidth]{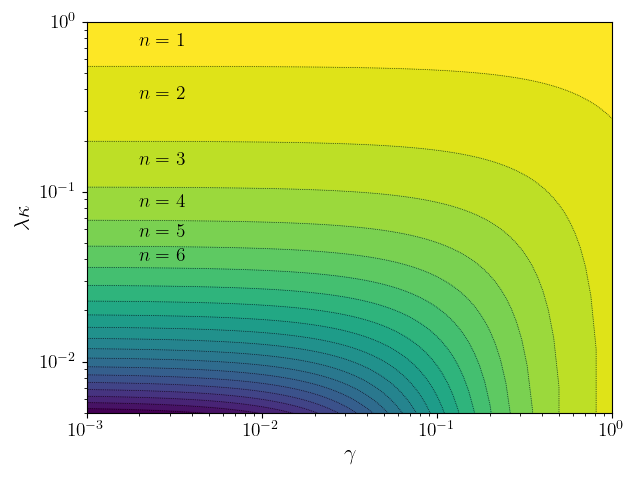}
		\includegraphics[width=0.49\linewidth]{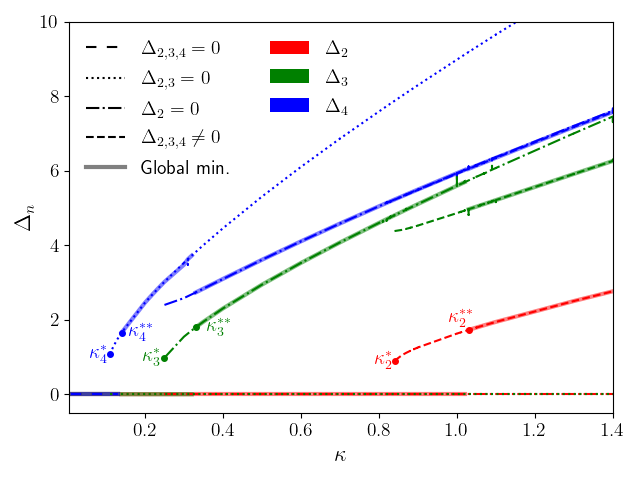}	
		\caption{Optimal launch strategies for non-convex SASP. $\underline{\text{Left}}$: In the case of an exponential decay, the best strategy is to launch $n$ searchers initially and none after. We show the corresponding optimal number of searchers as a function of $\kappa$ and $\gamma$. $\underline{\text{Right}}$: For the algebraic decay, finite optimal launch times are observed. We show all launch intervals $\Delta_n$ that are locally optimal as a function of $\kappa$ for $N=4$ and $\gamma = 0.01\lambda$. Different dashed line styles indicate different local minima while the solid line indicates the global one. Colors code for the second, third and fourth searcher.}	
		\label{fig:nonconvex}
	\end{center}
\end{figure*}

For the exponential test case $s_{exp}^{flat}$, we observe that all local minima are such that $t_i=0$ for $i\leq k$ and $t_i\to \infty$ for $i>k$: similarly to the mild exponential case, the optimal strategy is to launch a certain number of searchers initially and none after. However, this optimal number of initial searchers is not given by equation (\ref{eq:Nsim}) but can be found numerically by minimizing the mean search cost of the form $n\kappa + (1+n\gamma)\int_{0}^{\infty}s(t)^n dt$ with respect to $n$. We show the resulting optimal number of searchers to be launched in the $(\kappa, \gamma)$-plane in figure \ref{fig:nonconvex} and note that it does depend on $\gamma$, unlike the convex case.  

In the case of an algebraic decay, i.e. the test case $s_{alg}^{flat}$, we observe that the siuation is more complex as finite, non-zero launch times can be optimal. We show in figure \ref{fig:nonconvex} an example of the optimal strategies for $N=4$. For low values of $\kappa$, the only local minimum is located at $\Delta_{2,3,4}=0$: the optimal strategy is to launch all searchers simultaneously. As one progressively increases $\kappa$, a local minimum with $\Delta_4>0$ appears at $\kappa = \kappa_4^*$ but only becomes the global minimum for $\kappa=\kappa_4^{**}>\kappa_4^*$. Increasing again $\kappa$, a local minima with $\Delta_{2,3}>0$ appear at $\kappa_{2,3}^*$ and again become the global minima at $\kappa_{2,3}^{**}$. Because of the high number of local minima, the numerical optimization for a larger number of searchers becomes challenging. However, one can legitimately expect from our results in the convex case that the launch time of later searchers should be governed by the long-time behavior of the SASP, which is here algebraic. We thus conjecture that the time interval between consecutive launches should optimally reach a constant value, similarly to the convex SASP that decay algebraically.   

\section{Proof of principle}

\begin{figure*}[t]
	\includegraphics[width=\linewidth]{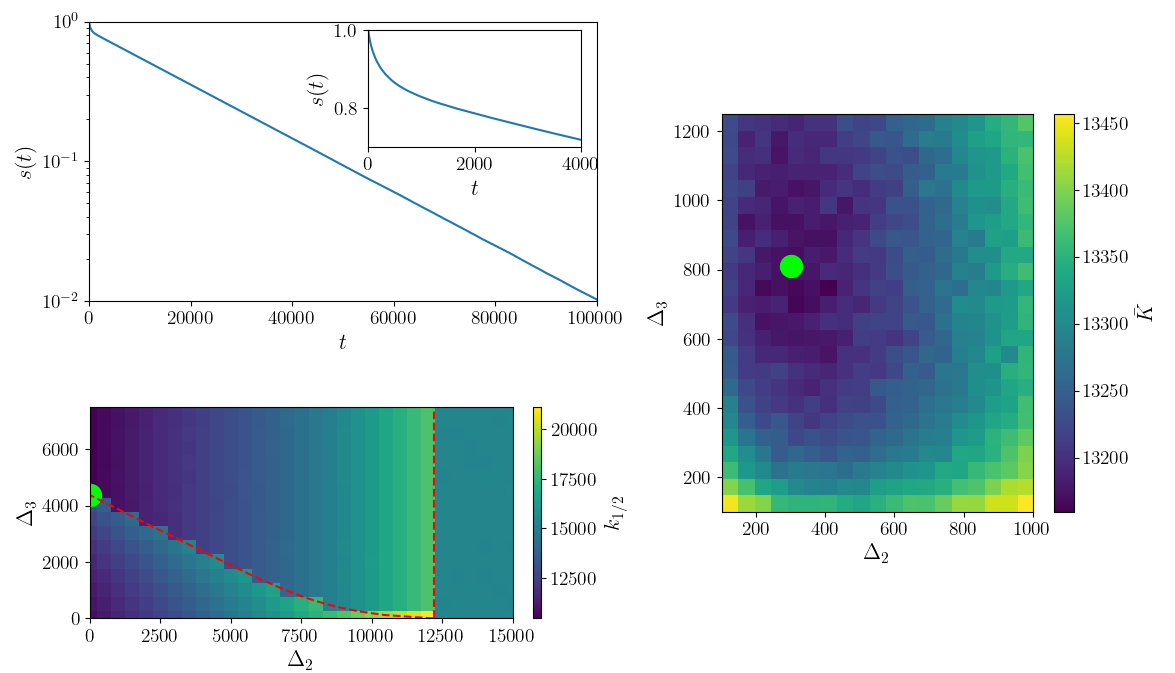}
	\caption{Numerical simulations of a Lévy flight process with $\alpha=5$ in which the target is placed at the origin of a box with periodic boundary conditions and searchers are launched from a position that is Gaussian distributed around this target. $\underline{\text{Top left}}$: SASP $s(t)$ as a function of $t$. The inset zooms on the short-time behavior. $\underline{\text{Bottom left}}$: Median search cost for a 3-searcher process as a function of $\Delta_2$ and $\Delta_3$ as sampled from the numerical simulations for $\kappa = 0.15\bar{T}_1$ and $\gamma = 0.01$. The dotted line are the boundary lines $\Lambda_1$ and $\Lambda_2$ as predicted by equation (\ref{eq:s_tauin}) while the green dot is the location of the minimum median cost as predicted by our theory, which coincides with the minimum obtained by numerical simulations. $\underline{\text{Right}}$: The same analysis is shown for the mean search cost. Again, the theoretical prediction for the minimum search cost matches the simulation data.}
	\label{fig:levy}
\end{figure*}

In order to verify the validity of our results, we now compare our prediction with data from numerical simulations in an example of search process. We consider a Lévy flight process, known to be a rather efficient class of random walks for target searches. In a 2-dimensional square box of size $L$ with periodic conditions, we place a target at the origin $\mathbf{r}_0$. Searchers are initialized at random positions $\mathbf{r}_i = (x_i,y_i)$ where both $x_i$ and $y_i$ are drawn from a Gaussian distribution centered in $\mathbf{r}_0$ and with standard deviation $\sigma_{x,y}$. At each time step, the searcher jumps from its position $\mathbf{r}_t$ to a new position $\mathbf{r}_{t+1} = \mathbf{r}_t + \Delta \mathbf{r}$ where $\Delta \mathbf{r} = l(\cos \theta, \sin \theta)$. $\theta$ is a random angle uniformly distributed between $0$ and $2\pi$ while $l$ is a random jump length drawn from a Pareto distribution $p_l(l) \propto l^{-\alpha}$ for $l>1$ and $\alpha>1$. The process stops whenever the searcher reaches a position $\mathbf{r}$ such that $|\mathbf{r}-\mathbf{r}_0|<1$, i.e. it has reached the vicinity of the target. 

We start by running simulations for a single searcher using $\alpha=5$, $L=100$ and $\sigma_x=\sigma_y=10$ and evaluate the SASP $\bar{s}(t)$, which we show in figure \ref{fig:levy}. At short times, it decays very sharply. As the probability for the searcher to be initialized very close to the target is high, many trajectories are such that the searcher finds the target almost immediately after being launched, and thus the probability for the target to be not reached yet after a certain time decreases fast initially. At long times, we observe an exponential decay, which is also consistent with most SASP of searches in closed domains. The corresponding single-search mean first passage time is found to be $\bar{T}_1 = 19423$.

Based on this observation, our results from the previous section suggest that only one searcher should be launched initially, and later searchers should be launched at finite times with increasing consecutive launch intervals that eventually diverge. As an example, for $\kappa = 0.15\bar{T}_1$ and $\gamma = 0.01$, the numerical optimization performed in the last section  predicts that the optimal launch strategy for multiple searchers is obtained for $\Delta_2 = 0.015\bar{T}_1$, $\Delta_3 = 0.041\bar{T}_1$ and $\Delta_4 \to \infty$. In order to verify this prediction, we ran simulations with 3 searchers in which the second and third are launched at various times and for each set of launch intervals $(\Delta_2,\Delta_3)$ we computed the resulting MFPT. We plot it in figure \ref{fig:levy} in the $(\Delta_2, \Delta_3)$-plane in which we also indicate the location of the predicted optimal strategy and we observe a very good agreement as this predicted point coincides with the minimum obtained via numerical simulations. 

Similarly, we compare the predicted optimal strategy to minimize the median cost (i.e. the quantile with $z=0.5$). For the same values of $\kappa$ and $\gamma$, our prediction is such that the second searcher should be launched at $t=0$ and the third after $t=0.22\bar{T}_1$, knowing that any later time does not impact the median cost. This prediction matches perfectly the simulation data, which also coincide very well with the predicted domain lines $\Lambda_n$ defined and described in section III, as shown in figure \ref{fig:levy}.

\section{Comparison with stochastic resetting}

The process of launching a new searcher until a target has been found is reminiscent of search processes with stochastic resetting. Instead of launching new searchers at the initial position in intervals one could instead reset the first searcher to its initial position at a certain rate. Such processes have attracted a lot of interest recently, both on the theoretical and experimental sides \cite{pal2015diffusion, evans2020stochastic, tal2020experimental, sandev2022heterogeneous, gupta2022stochastic}.
Fine-tuning the time at which the searcher is subject to resetting can have a significant impact on the overall search efficiency \cite{pal2019first, ahmad2019first, chen2022first, linn2023first}. Given the cost for resetting a searcher and sustaining it over time, we can legitimately wonder whether resetting the first searcher is more cost-effective than launching a new one. While the time it takes to find the target will be reduced by launching a new searcher, the cost for sustaining multiple agents rather than a single one may overcompensate the gain in the overall search time. 

Let us first define the search cost $K_r$ of a resetting process in accordance with our search cost for the launch process, reading
\begin{equation}
K_r= (1+\gamma_r) T + \kappa_r N_r
\end{equation}
where $T$ is the first-passage time and $N_r$ is the number of resetting events. Here $\gamma_r = \gamma$ is the sustaining rate of the searcher, and $\kappa_r$ may vary from one trajectory to the next depending on the process. There exists e.g. processes where the cost for resetting a searcher depend on the distance that it needs to be displaced in order to place it back to its initial position. While it is unlikely to derive general relevant results for the comparison between launching and resetting strategies, we will treat a canonical example in the next lines, namely the one-dimensional diffusive search, that the reader could then adapt to their favourite random search process. 

We consider a one-dimensional infinite system. A target is placed a position $x_T>0$, searchers are initialized at $x=0$ and evolve according to standard diffusive Brownian motion with diffusion coefficient $D$. In this case, it is well-known that the SASP is found to be 
\begin{equation}
\bar{s}(t) = \text{erf}\left(x_T/2\sqrt{Dt}\right)
\end{equation}
When it comes to resetting the first searcher to its initial position, two resetting costs have been considered in the literature. Either the resetting is proportional to the work one needs to perform to reset which typically scales like the distance it has reached just before being reset, or it is constant. In reference \cite{chechkin2018random}, Chechkin et al. showed that in a resetting process where the time between two resetting events was not conditioned on anything and was drawn from a single probability distribution, the optimal strategy is to make this distribution as narrow as possible around a mean value $\Delta$. Following the same methodology, we can show that the mean search cost $\bar{K}_r$ is minimized for a fixed resetting time interval $\Delta$, in which case it reads
\begin{equation}
\bar{K}_r (\Delta) = \frac{(1+\gamma_r) G(\Delta) + \bar{\kappa_r}(\Delta)}{F(\Delta)}
\end{equation}
where $F(\Delta) = 1-s(\Delta)$ and $G(\Delta) = \int_{0}^{\Delta} s(\tau) d\tau$ and $\bar{\kappa}_r$ is the average of the resetting costs over all possible resetting events. In the case of a resetting cost proportional to the distance $|\Delta x|$ between the searcher and its initial position upon resetting
as used e.g. in \cite{sunil2023cost, talfriedman2024smart}, we obtain
\begin{equation}
\bar{\kappa}_r(\Delta) = f\frac{\int_{-\infty}^{x_T} dx |x| c(x,\Delta)}{\int_{-\infty}^{x_T} dx c(x,\Delta)}
\end{equation}
where $f$ is the cost per unit length and $c(x,t)$ is the particle density in an experiment with absorbing boundary at $x=x_T$, i.e. $\int_{-\infty}^{x_T}c(x,t) dx = \bar{s}(t)$. In particular for the 1D diffusion model, $c(x,t)$ is known to be  \cite{evans2011diffusion}
\begin{equation}
c(x,t) = \frac{1}{\sqrt{4\pi D t}}\left(e^{-\frac{x^2}{4Dt}} - e^{-\frac{(x-2x_T)^2}{4Dt}} \right)
\end{equation}
which allows to then determine $\bar{s}(t)$, $F(t)$, $G(t)$ and eventualy 
\begin{align}
\bar{\kappa}_r (\Delta) = f l_d &\left[ \frac{2}{\sqrt{\pi}}\left( 1- e^{-\eta^2} \right) \right. \nonumber \\ 
&\left. + \eta \left( 1 - 2\text{erf}\left(\eta\right) + \text{erf}\left(\frac{\eta}{2} \right) \right)\right] 
\end{align}
where we have defined $l_d = \sqrt{D\Delta}$ and $\eta = x_T/l_d$, which is equivalent to $f x_T$ as $\Delta \to \infty$. For a constant resetting cost, we simply have $\bar{\kappa}_r(\Delta) = \kappa_r^c$.

\begin{figure}
	\begin{center}
		\includegraphics[width=\linewidth]{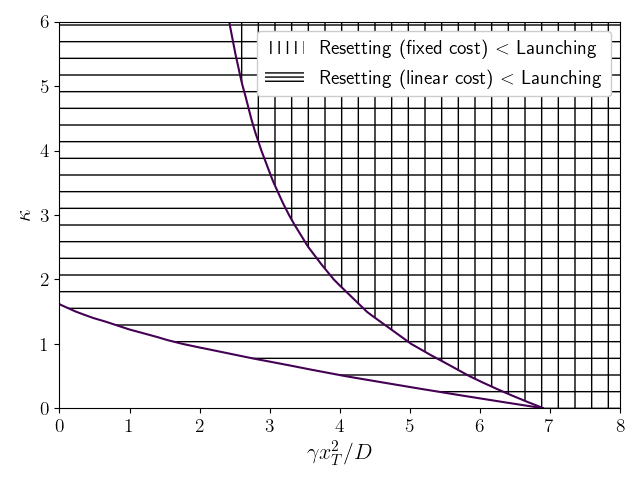}
		\caption{Preferential strategy (resetting or launching new agents) for the 1-dimensional diffusive search in the $(\kappa,\gamma)$-plane, where we have used $\kappa = \kappa_r = f_r x_T$.}
		\label{fig:resetting_vs_hiring} 
	\end{center}
\end{figure}

For both constant and linear resetting costs, we can minimize the total search cost $\bar{K}_r$ with respect to $\Delta$ in order to find the optimal resetting time. We can then compare the corresponding optimal search cost with the optimal cost that one would find if one would launch new searchers at a constant time interval instead of resetting the first one. This optimal search cost in the launch process can be found by setting $t_{i} = (i-1)\Delta$, calculate the corresponding search cost and minimize it with respect to $\Delta$. Figure \ref{fig:resetting_vs_hiring} shows for which values of $\kappa$ and $\gamma$ the optimal cost in the launch process is lower than the optimal search cost in the resetting process. To do this comparison, we need to set $\kappa_r^c$ and $f_r$ to a value which we choose as $\kappa = \kappa_r^c = f_rx_T$. It appears that launching new searchers is preferable to resetting the first one as long as the costs for sustaining a searcher, $\gamma$ multiplied with the diffusive timescale, $x_T^2/D$, is sufficiently small, provided the 
costs for creating new searchers, $\kappa$ is also small enough. For linear resetting cost, there is even a value of $\kappa$ above which  resetting is always better than launching regardless of the value of $\gamma$.

Although we cannot draw general quantitative conclusions about the efficiency of launching compared to resetting from this single example since it is based on a specific model, it seems reasonable to assume that if $\kappa$ and $\gamma$ are not too large, the gain in the overall search time will dominate the costs of launching and sustaining new searchers. Note, however, that in practical applications, it could be much harder, if not unfeasible, to reset searchers, in contrast to adding new searchers.

\section{Discussion and conclusion}

In this manuscript, we have discussed the optimal strategies for when to launch non-interacting agents in a collective search process. We have defined a search cost that accounts for the overall search time, the amount of resources used by the searchers as they search and the cost for launching a searcher. We have quantified when one should launch consecutive searchers in order to minimize the search costs, either its mean or its $q$-quantiles -- that is the cost value for which the probability of observing lower costs is $q^{-1}$ -- and described it in terms of the single-agent survival probability of the target, i.e. the probability that the target has not ben found after a certain time in a process with just one searcher. For both the mean and the quantiles, we have shown that if the SASP is logarithmically convex, there exists a unique local minimum of the search cost with respect to the launch times. Regarding the optimization of the quantiles, we have derived analytically the optimal launch strategy to minimize the probability for the search cost to be larger than a certain fixed value, which consists in launching an optimized number of agents initially and the next ones later than a certain time defined in terms of the SASP. We have also shown that this optimal strategy does not fix the launch times of later agents such that they can be adjusted to minimize subsequent quantiles. Regarding the mean search cost, we have shown that for situations where only one local minimum exists, the only instance where multiple searchers should be introduced simultaneously is at the start of the process and their optimal number is entirely governed by the launch cost and the initial slope of the SASP. Then, the details of the entire launching strategy was analyzed on a variety of test cases combining short and long-time behaviors of the SASP. In particular, we have shown that the steeper the initial decay of the SASP is, the longer the time interval between the early searchers. However, the optimal time interval between launches of late searchers diverge if the SASP decays exponentially while it reaches a constant value for algebraically decaying SASP. Finally, we have compared our results with search process subject to stochastic resetting. Using the example of the one-dimensional diffusive search, we showed that launching searchers is preferable to resetting the first one to its initial position as long as the launch cost and sustaining cost are not too large. 

Our work is the first to investigate in details the optimization of a search process with respect to its number of searchers. The results exposed apply to a wide variety of processes but it must not be forgotten that some important assumptions were made along the way. In addition to the absence of interactions between agents, we also assumed that the SASP decays to zero, implying that each agent would eventually find the target if it would search for long enough. This is e.g. not the case in diffusive searches in open space in dimension 2 or more, and it would be interesting to check to which extent the optimal strategies identified in this paper perform for such cases. 

While the timing for the launches matters a lot in the overall search cost, as we have shown in this work, one might also wonder about the locations of the launches. Is there an optimal way to select initial positions of each searcher in order to minimize the search cost ? This question is again reminiscent of stochastic resetting but in the context of collective searchers and should undoubtedly be treated in depth in the future. The analogy with recent studies on stochastic resetting can be stretched even further as one could also construct processes where the launching of new agents is conditioned on the evolution of external quantities, such as the proportion of the search domain that has been scanned at a certain time, or the current position of the agents already launched. This type of questions enter the realm of decision making, a cross-disciplinary research field that can be used to substantially increase search efficiency and is a perfect playground for developing machine-learning algorithms. We believe that our work is a step in this direction and will motivate future studies to develop more and more refined collective search strategies.

\section*{Acknowledgements} We acknowledge financial support by the DFG via the Collaborative Research Center SFB 1027. This work was supported by the Medical Research Council (Grant No. MR/Y003845/1).

\bibliographystyle{unsrt}

\begin{thebibliography}{10}
	
	\bibitem{grebenkov2024target}
	Denis Grebenkov, Ralf Metzler, and Gleb Oshanin.
	\newblock Target search problems.
	\newblock In {\em Target Search Problems}, pages 1--29. Springer, 2024.
	
	\bibitem{collins1949diffusion}
	Frank~C Collins and George~E Kimball.
	\newblock Diffusion-controlled reaction rates.
	\newblock {\em Journal of colloid science}, 4(4):425--437, 1949.
	
	\bibitem{rice1985diffusion}
	SA~Rice.
	\newblock {\em Diffusion-Limited Reactions}.
	\newblock Elsevier, 1985.
	
	\bibitem{kolomeisky2011physics}
	Anatoly~B Kolomeisky.
	\newblock Physics of protein--dna interactions: mechanisms of facilitated
	target search.
	\newblock {\em Physical Chemistry Chemical Physics}, 13(6):2088--2095, 2011.
	
	\bibitem{harlin2009chemokine}
	Helena Harlin, Yuru Meng, Amy~C Peterson, Yuanyuan Zha, Maria Tretiakova, Craig
	Slingluff, Mark McKee, and Thomas~F Gajewski.
	\newblock Chemokine expression in melanoma metastases associated with cd8+
	t-cell recruitment.
	\newblock {\em Cancer research}, 69(7):3077--3085, 2009.
	
	\bibitem{krummel2016t}
	Matthew~F Krummel, Frederic Bartumeus, and Audrey G{\'e}rard.
	\newblock T cell migration, search strategies and mechanisms.
	\newblock {\em Nature Reviews Immunology}, 16(3):193--201, 2016.
	
	\bibitem{galeano2020cytotoxic}
	Jorge~Luis Galeano~Ni{\~n}o, Sophie~V Pageon, Szun~S Tay, Feyza Colakoglu,
	Daryan Kempe, Jack Hywood, Jessica~K Mazalo, James Cremasco, Matt~A Govendir,
	Laura~F Dagley, et~al.
	\newblock Cytotoxic t cells swarm by homotypic chemokine signalling.
	\newblock {\em Elife}, 9:e56554, 2020.
	
	\bibitem{hassell1978foraging}
	MP~Hassell and TRE Southwood.
	\newblock Foraging strategies of insects.
	\newblock {\em Annual review of ecology and systematics}, 9:75--98, 1978.
	
	\bibitem{bartumeus2009optimal}
	Frederic Bartumeus and Jordi Catalan.
	\newblock Optimal search behavior and classic foraging theory.
	\newblock {\em Journal of Physics A: Mathematical and Theoretical},
	42(43):434002, 2009.
	
	\bibitem{sumpter2010collective}
	David~JT Sumpter.
	\newblock {\em Collective animal behavior}.
	\newblock Princeton University Press, 2010.
	
	\bibitem{viswanathan2011physics}
	Gandhimohan~M Viswanathan, Marcos~GE Da~Luz, Ernesto~P Raposo, and H~Eugene
	Stanley.
	\newblock {\em The physics of foraging: an introduction to random searches and
		biological encounters}.
	\newblock Cambridge University Press, 2011.
	
	\bibitem{pyke2019animal}
	Graham Pyke.
	\newblock Animal movements: an optimal foraging approach.
	\newblock In {\em Encyclopedia of animal behavior}, pages 149--156. Elsevier
	Academic Press, 2019.
	
	\bibitem{senanayake2016search}
	Madhubhashi Senanayake, Ilankaikone Senthooran, Jan~Carlo Barca, Hoam Chung,
	Joarder Kamruzzaman, and Manzur Murshed.
	\newblock Search and tracking algorithms for swarms of robots: A survey.
	\newblock {\em Robotics and Autonomous Systems}, 75:422--434, 2016.
	
	\bibitem{drew2021multi}
	Daniel~S Drew.
	\newblock Multi-agent systems for search and rescue applications.
	\newblock {\em Current Robotics Reports}, 2:189--200, 2021.
	
	\bibitem{viswanathan1996levy}
	Gandhimohan~M Viswanathan, Vsevolod Afanasyev, Sergey~V Buldyrev, Eugene~J
	Murphy, Peter~A Prince, and H~Eugene Stanley.
	\newblock L{\'e}vy flight search patterns of wandering albatrosses.
	\newblock {\em Nature}, 381(6581):413--415, 1996.
	
	\bibitem{viswanathan1999optimizing}
	Gandimohan~M Viswanathan, Sergey~V Buldyrev, Shlomo Havlin, MGE Da~Luz,
	EP~Raposo, and H~Eugene Stanley.
	\newblock Optimizing the success of random searches.
	\newblock {\em nature}, 401(6756):911--914, 1999.
	
	\bibitem{oshanin2007intermittent}
	G~Oshanin, HS~Wio, K~Lindenberg, and SF~Burlatsky.
	\newblock Intermittent random walks for an optimal search strategy:
	one-dimensional case.
	\newblock {\em Journal of Physics: Condensed Matter}, 19(6):065142, 2007.
	
	\bibitem{yang2009cuckoo}
	Xin-She Yang and Suash Deb.
	\newblock Cuckoo search via l{\'e}vy flights.
	\newblock In {\em 2009 World congress on nature \& biologically inspired
		computing (NaBIC)}, pages 210--214. Ieee, 2009.
	
	\bibitem{reynolds2009levy}
	Andy~M Reynolds and Christopher~J Rhodes.
	\newblock The l{\'e}vy flight paradigm: random search patterns and mechanisms.
	\newblock {\em Ecology}, 90(4):877--887, 2009.
	
	\bibitem{benichou2011intermittent}
	Olivier B{\'e}nichou, Claude Loverdo, Michel Moreau, and Raphael Voituriez.
	\newblock Intermittent search strategies.
	\newblock {\em Reviews of Modern Physics}, 83(1):81, 2011.
	
	\bibitem{kusmierz2014first}
	Lukasz Kusmierz, Satya~N Majumdar, Sanjib Sabhapandit, and Gr{\'e}gory Schehr.
	\newblock First order transition for the optimal search time of l{\'e}vy
	flights with resetting.
	\newblock {\em Physical review letters}, 113(22):220602, 2014.
	
	\bibitem{chechkin2018random}
	A~Chechkin and IM3845437 Sokolov.
	\newblock Random search with resetting: a unified renewal approach.
	\newblock {\em Physical review letters}, 121(5):050601, 2018.
	
	\bibitem{bressloff2020search}
	Paul~C Bressloff.
	\newblock Search processes with stochastic resetting and multiple targets.
	\newblock {\em Physical Review E}, 102(2):022115, 2020.
	
	\bibitem{evans2020stochastic}
	Martin~R Evans, Satya~N Majumdar, and Gr{\'e}gory Schehr.
	\newblock Stochastic resetting and applications.
	\newblock {\em Journal of Physics A: Mathematical and Theoretical},
	53(19):193001, 2020.
	
	\bibitem{pal2020search}
	Arnab Pal, {\L}ukasz Ku{\'s}mierz, and Shlomi Reuveni.
	\newblock Search with home returns provides advantage under high uncertainty.
	\newblock {\em Physical Review Research}, 2(4):043174, 2020.
	
	\bibitem{tejedor2012optimizing}
	Vincent Tejedor, Raphael Voituriez, and Olivier B{\'e}nichou.
	\newblock Optimizing persistent random searches.
	\newblock {\em Physical review letters}, 108(8):088103, 2012.
	
	\bibitem{meyer2021optimal}
	Hugues Meyer and Heiko Rieger.
	\newblock Optimal non-markovian search strategies with n-step memory.
	\newblock {\em Physical Review Letters}, 127(7):070601, 2021.
	
	\bibitem{barbier2022self}
	A~Barbier-Chebbah, O~B{\'e}nichou, and R~Voituriez.
	\newblock Self-interacting random walks: Aging, exploration, and first-passage
	times.
	\newblock {\em Physical Review X}, 12(1):011052, 2022.
	
	\bibitem{klimek2022optimal}
	Anton Klimek and Roland~R Netz.
	\newblock Optimal non-markovian composite search algorithms for spatially
	correlated targets.
	\newblock {\em Europhysics Letters}, 139(3):32003, 2022.
	
	\bibitem{altshuler2023environmental}
	Amy Altshuler, Ofek~Lauber Bonomo, Nicole Gorohovsky, Shany Marchini, Eran
	Rosen, Ofir Tal-Friedman, Shlomi Reuveni, and Yael Roichman.
	\newblock Environmental memory facilitates search with home returns.
	\newblock {\em arXiv preprint arXiv:2306.12126}, 2023.
	
	\bibitem{romanczuk2009collective}
	Pawel Romanczuk, Iain~D Couzin, and Lutz Schimansky-Geier.
	\newblock Collective motion due to individual escape and pursuit response.
	\newblock {\em Physical Review Letters}, 102(1):010602, 2009.
	
	\bibitem{martinez2013optimizing}
	Ricardo Mart{\'\i}nez-Garc{\'\i}a, Justin~M Calabrese, Thomas Mueller, Kirk~A
	Olson, and Crist{\'o}bal L{\'o}pez.
	\newblock Optimizing the search for resources by sharing information:<?
	format?> mongolian gazelles as a case study.
	\newblock {\em Physical Review Letters}, 110(24):248106, 2013.
	
	\bibitem{janosov2017group}
	Mil{\'a}n Janosov, Csaba Vir{\'a}gh, G{\'a}bor V{\'a}s{\'a}rhelyi, and
	Tam{\'a}s Vicsek.
	\newblock Group chasing tactics: how to catch a faster prey.
	\newblock {\em New Journal of Physics}, 19(5):053003, 2017.
	
	\bibitem{kamimura2019group}
	Atsushi Kamimura and Toru Ohira.
	\newblock {\em Group Chase and Escape: Fusion of Pursuits-Escapes and
		Collective Motions}.
	\newblock Springer Nature, 2019.
	
	\bibitem{surendran2019spatial}
	Anudeep Surendran, Michael~J Plank, and Matthew~J Simpson.
	\newblock Spatial structure arising from chase-escape interactions with
	crowding.
	\newblock {\em Scientific reports}, 9(1):14988, 2019.
	
	\bibitem{mejia2011first}
	Carlos Mej{\'\i}a-Monasterio, Gleb Oshanin, and Gr{\'e}gory Schehr.
	\newblock First passages for a search by a swarm of independent random
	searchers.
	\newblock {\em Journal of Statistical Mechanics: Theory and Experiment},
	2011(06):P06022, 2011.
	
	\bibitem{bernardi2022run}
	Davide Bernardi and Benjamin Lindner.
	\newblock Run with the brownian hare, hunt with the deterministic hounds.
	\newblock {\em Physical Review Letters}, 128(4):040601, 2022.
	
	\bibitem{biroli2023critical}
	Marco Biroli, Satya~N Majumdar, and Gr{\'e}gory Schehr.
	\newblock Critical number of walkers for diffusive search processes with
	resetting.
	\newblock {\em Physical Review E}, 107(6):064141, 2023.
	
	\bibitem{tani2014optimal}
	Noriyuki Tani, David Quint, and Ajay Gopinathan.
	\newblock Optimal cooperative searching using purely repulsive interactions.
	\newblock {\em Biophysical Journal}, 106(2):788a, 2014.
	
	\bibitem{ro2023target}
	Sunghan Ro, Juyeon Yi, and Yong~Woon Kim.
	\newblock Target searches of interacting brownian particles in dilute systems.
	\newblock {\em Physical Review E}, 107(6):064143, 2023.
	
	\bibitem{meyer2024collective}
	Hugues Meyer, Adam Wysocki, and Heiko Rieger.
	\newblock Collective chemotactic search strategies, 2024.
	
	\bibitem{mattos2012first}
	Thiago~G Mattos, Carlos Mej{\'\i}a-Monasterio, Ralf Metzler, and Gleb Oshanin.
	\newblock First passages in bounded domains: When is the mean first passage
	time meaningful?
	\newblock {\em Physical Review E—Statistical, Nonlinear, and Soft Matter
		Physics}, 86(3):031143, 2012.
	
	\bibitem{shaebani2018unraveling}
	M.~Reza Shaebani, Robin Jose, Christian Sand, and Ludger Santen.
	\newblock Unraveling the structure of treelike networks from first-passage
	times of lazy random walkers.
	\newblock {\em Phys. Rev. E}, 98:042315, Oct 2018.
	
	\bibitem{bray2013persistence}
	Alan~J Bray, Satya~N Majumdar, and Gr{\'e}gory Schehr.
	\newblock Persistence and first-passage properties in nonequilibrium systems.
	\newblock {\em Advances in Physics}, 62(3):225--361, 2013.
	
	\bibitem{godec2016first}
	Alja{\v{z}} Godec and Ralf Metzler.
	\newblock First passage time distribution in heterogeneity controlled kinetics:
	going beyond the mean first passage time.
	\newblock {\em Scientific reports}, 6(1):20349, 2016.
	
	\bibitem{redner2023first}
	Sidney Redner.
	\newblock A first look at first-passage processes.
	\newblock {\em Physica A: Statistical Mechanics and its Applications},
	631:128545, 2023.
	
	\bibitem{koren2007leapover}
	Tal Koren, Michael~A Lomholt, Aleksei~V Chechkin, Joseph Klafter, and Ralf
	Metzler.
	\newblock Leapover lengths and first passage time statistics for l{\'e}vy
	flights.
	\newblock {\em Physical review letters}, 99(16):160602, 2007.
	
	\bibitem{slepian1962one}
	David Slepian.
	\newblock The one-sided barrier problem for gaussian noise.
	\newblock {\em Bell System Technical Journal}, 41(2):463--501, 1962.
	
	\bibitem{pal2015diffusion}
	Arnab Pal.
	\newblock Diffusion in a potential landscape with stochastic resetting.
	\newblock {\em Physical Review E}, 91(1):012113, 2015.
	
	\bibitem{tal2020experimental}
	Ofir Tal-Friedman, Arnab Pal, Amandeep Sekhon, Shlomi Reuveni, and Yael
	Roichman.
	\newblock Experimental realization of diffusion with stochastic resetting.
	\newblock {\em The journal of physical chemistry letters}, 11(17):7350--7355,
	2020.
	
	\bibitem{sandev2022heterogeneous}
	Trifce Sandev, Viktor Domazetoski, Ljupco Kocarev, Ralf Metzler, and Aleksei
	Chechkin.
	\newblock Heterogeneous diffusion with stochastic resetting.
	\newblock {\em Journal of Physics A: Mathematical and Theoretical},
	55(7):074003, 2022.
	
	\bibitem{gupta2022stochastic}
	Shamik Gupta and Arun~M Jayannavar.
	\newblock Stochastic resetting: A (very) brief review.
	\newblock {\em Frontiers in Physics}, 10:789097, 2022.
	
	\bibitem{pal2019first}
	Arnab Pal and VV~Prasad.
	\newblock First passage under stochastic resetting in an interval.
	\newblock {\em Physical Review E}, 99(3):032123, 2019.
	
	\bibitem{ahmad2019first}
	Saeed Ahmad, Indrani Nayak, Ajay Bansal, Amitabha Nandi, and Dibyendu Das.
	\newblock First passage of a particle in a potential under stochastic
	resetting: A vanishing transition of optimal resetting rate.
	\newblock {\em Physical Review E}, 99(2):022130, 2019.
	
	\bibitem{chen2022first}
	Hanshuang Chen and Feng Huang.
	\newblock First passage of a diffusing particle under stochastic resetting in
	bounded domains with spherical symmetry.
	\newblock {\em Physical Review E}, 105(3):034109, 2022.
	
	\bibitem{linn2023first}
	Samantha Linn and Sean~D Lawley.
	\newblock First-passage times under frequent stochastic resetting.
	\newblock {\em Physical Review E}, 108(2):024114, 2023.
	
	\bibitem{sunil2023cost}
	John~C Sunil, Richard~A Blythe, Martin~R Evans, and Satya~N Majumdar.
	\newblock The cost of stochastic resetting.
	\newblock {\em Journal of Physics A: Mathematical and Theoretical},
	56(39):395001, 2023.
	
	\bibitem{talfriedman2024smart}
	Ofir Tal-Friedman, Tommer~D. Keidar, Shlomi Reuveni, and Yael Roichman.
	\newblock Smart resetting: An energy-efficient strategy for stochastic search
	processes, 2024.
	
	\bibitem{evans2011diffusion}
	Martin~R. Evans and Satya~N. Majumdar.
	\newblock Diffusion with stochastic resetting.
	\newblock {\em Phys. Rev. Lett.}, 106:160601, Apr 2011.
	
\end{thebibliography}

\end{document}